\documentclass{jpp}

\usepackage{graphicx}% Include figure files
\usepackage{caption}
\usepackage{subcaption}
\usepackage{dcolumn}% Align table columns on decimal point
\usepackage{bm}% bold math
\usepackage[dvipsnames]{xcolor}
\usepackage{amssymb}
\usepackage{amsmath}
\usepackage{verbatim}
\usepackage{appendix}
\usepackage{listings}
\usepackage{units}
\usepackage{csquotes}

\usepackage{hyperref}

\usepackage{etoolbox}
\makeatletter
\patchcmd{\hyper@makecurrent}{%
    \ifx\Hy@param\Hy@chapterstring
        \let\Hy@param\Hy@chapapp
    \fi
}{%
    \iftoggle{inappendix}{%true-branch
        \@checkappendixparam{chapter}%
        \@checkappendixparam{section}%
        \@checkappendixparam{subsection}%
        \@checkappendixparam{subsubsection}%
        \@checkappendixparam{paragraph}%
        \@checkappendixparam{subparagraph}%
    }{}%
}{}{\errmessage{failed to patch}}

\newcommand*{\@checkappendixparam}[1]{%
    \def\@checkappendixparamtmp{#1}%
    \ifx\Hy@param\@checkappendixparamtmp
        \let\Hy@param\Hy@appendixstring
    \fi
}
\makeatletter

\newtoggle{inappendix}
\togglefalse{inappendix}

\apptocmd{\appendix}{\toggletrue{inappendix}}{}{\errmessage{failed to patch}}
\apptocmd{\subappendices}{\toggletrue{inappendix}}{}{\errmessage{failed to patch}}
\newcommand*\obar[2][0.75]{% OverBAR, adds bar over an element
    \sbox{\myboxA}{$\m@th#2$}%
    \setbox\myboxB\null% Phantom box
    \ht\myboxB=\ht\myboxA%
    \dp\myboxB=\dp\myboxA%
    \wd\myboxB=#1\wd\myboxA% Scale phantom
    \sbox\myboxB{$\m@th\overline{\copy\myboxB}$}%  Overlined phantom
    \setlength\mylenA{\the\wd\myboxA}%   calc width diff
    \addtolength\mylenA{-\the\wd\myboxB}%
    \ifdim\wd\myboxB<\wd\myboxA%
       \rlap{\hskip 0.5\mylenA\usebox\myboxB}{\usebox\myboxA}%
    \else
        \hskip -0.5\mylenA\rlap{\usebox\myboxA}{\hskip 0.5\mylenA\usebox\myboxB}%
    \fi}
\makeatother

\definecolor{colorexample}{RGB}{250,143,56}

\newcommand{\ordo}[1]{{\cal O}\left( #1 \right)}

\DeclareMathOperator\erf{erf}
\renewcommand{\vec}[1]{\boldsymbol{#1}}

\newcommand{\e}{\ensuremath{\mathrm{e}}}
\newcommand{\E}{\ensuremath{\mathcal{E}}}
\newcommand{\w}{\ensuremath{w}}
\newcommand{\A}{\ensuremath{\tilde{\Phi}(B_0)}}
\newcommand{\C}{\ensuremath{X}}

\renewcommand{\d}{\ensuremath{\mathrm{d}}}

\newcommand{\remove}[1]{{}}

\newcommand*{\nouncite}[1]{\citet{#1}}
\newcommand{\lang}{\left\langle}
\newcommand{\rang}{\right\rangle}

\usepackage{stackengine}
\stackMath

\newcommand{\appref}[1]{\hyperref[#1]{Appendix~\ref{#1}}}

\usepackage{morefloats}
\begin{document}

% \preprint{AIP/123-QED}

\shorttitle{Flux-surface impurity density variation and impurity transport}
\shortauthor{S.\ Buller, H.M.\ Smith, P.\ Helander, A.\ Moll\'{e}n, S.L.\ Newton, I.\ Pusztai}

\title{Collisional transport of impurities with flux-surface varying density in Stellarators}
\author{S.\ Buller\aff{1} \corresp{\email{bstefan@chalmers.se}}, H.M.\ Smith\aff{2}, P.\ Helander\aff{2}, A.\ Moll\'{e}n\aff{2}, S.L.\ Newton\aff{3}, I.\ Pusztai\aff{1}} 
\affiliation{\aff{1}Department of Physics, Chalmers University of Technology,
  SE-41296 G\"{o}teborg, Sweden
  \aff{2}Max-Planck-Institut f\"{u}r Plasmaphysik, 17491 Greifswald, Germany
  \aff{3}CCFE, Culham Science Centre, Abingdon, Oxon OX14 3DB, UK
}

%\author{S.L. Newton}
%\affiliation{Department of Physics, Chalmers University of Technology,
%  SE-41296 G\"{o}teborg, Sweden}
%\affiliation{CCFE, Culham Science Centre, Abingdon, Oxon OX14 3DB, UK}
%\author{J.T. Omotani}
%\affiliation{Department of Physics, Chalmers University of Technology,
%  SE-41296 G\"{o}teborg, Sweden}
%\author{M. Landreman}
%\affiliation{Insitute for Research in Electronics and  Applied Physics,
%  University of Maryland, College Park, MD 20742, USA.  }
%\author{T. F\"{u}l\"{o}p}
%\affiliation{Department of Physics, Chalmers University of Technology,
%  SE-41296 G\"{o}teborg, Sweden} 


%\pacs{ }% PACS, the Physics and Astronomy
%\keywords{classical transport, neoclassical transport, stellarator, impurity}

\maketitle
 \begin{abstract}
High-$Z$ impurities in magnetic confinement devices are prone to develop density variations on the flux-surface, which can significantly affect their transport. In this paper, we generalize earlier analytic stellarator calculations of the neoclassical radial impurity flux in the mixed-collisionality regime (collisional impurities and low-collisionality bulk ions) to include the effect of such flux-surface variations. 
We find that only in the homogeneous density case is the transport of highly collisional impurities (in the Pfirsch-Schl\"{u}ter regime) independent of the radial electric field. 
We study these effects for a Wendelstein 7-X (W7-X) vacuum field% and a 4-Fourier-component approximation of a Large Helical Device (LHD) magnetic field [Beidler et al. Nucl. Fusion 51, 076001 (2011)]\todo{Can't calculate classical with the file I have}
, with simple analytic models for the potential perturbation, under the assumption that the impurity density is given by a Boltzmann response to a perturbed potential. In the W7-X case studied, we find that larger amplitude potential perturbations cause the radial electric field to dominate the transport of the impurities. %In LHD, however, the amplitude of the perturbation must be larger to have a large effect on the transport due to the radial electric field.
In addition, we find that classical impurity transport can be larger than the neoclassical transport in W7-X. %\todo{Can't say for LHD.}
\end{abstract}

%-----------------------------------------------------------
\section{Introduction}
\label{sec:intro}
At fusion-relevant temperatures, heavy impurities in high ionisation states, ``high-$Z$ impurities'', emit a significant amount of radiation, and even a tiny fraction of impurity ions radiate enough power to seriously challenge the power balance in a reactor. High-$Z$ impurities thus cannot be allowed to accumulate in the center of a magnetic-confinement fusion reactor. 

In tokamaks, impurities are expelled from the core of the reactor by neoclassical transport if their temperature gradient is sufficiently large -- a phenomenon known as \emph{temperature screening}. In stellarators, the outlook has been more pessimistic, as the radial transport is not independent of the radial electric field, and an inward pointing electric field is predicted for a stellarator reactor, which would transport impurities inwards \citep{hirschW7AS2008}.

However, recent analytical results on neoclassical stellarator impurity transport have shown that when the plasma is in a mixed-collisionality regime -- where the bulk ions are at low collisionality ($1/\nu$ or $\sqrt{\nu}$ regimes) and the impurity ions are collisional -- the radial impurity flux becomes independent of the electric field, which allows temperature screening to be effective in stellarators \citep{newton2017,helanderPRL2017}. This is due to a cancellation between the flux driven by impurity parallel flow and the ion thermodynamic forces. A similar cancellation is also found in the regimes where both ions and impurities are collisional \citep{braun2010a}, although in this case, thermodiffusion is usually inward unless the effective charge is very small, so no temperature screening occurs \citep{rutherford1974}.

Additionally, high-$Z$ impurities are sensitive to flux-surface variations in the electrostatic potential, in response to which they can develop density variations on flux-surfaces.
Such variations can have large effects on the neoclassical transport, as has been demonstrated analytically \citep{angioni2014,calvo2018} and numerically \citep{angioniWJET2014,garcia2017,mollen2018} for tokamaks and stellarators. Turbulent transport is also know to be affected by these variations, see for example \citet{mollen2012,mollen2014,angioniWJET2014}.
% could cite "Pfirsch–Schl\"uter impurity transport in stellarator edge plasmas with large radial gradients" here; it contains a mistake and misses the ExB flux, but supports that Er affects transport when variations are taken into account.

In this work, we generalize the analytical calculation in \citep{newton2017} to account for flux-surface variation of the impurity density in stellarators, using a fluid description for the impurities and solving for the ion distribution function in the $1/\nu$ regime. Our expression for the impurity flux agrees with that in \citet{calvo2018}, where the same problem is treated fully kinetically. Like \citet{calvo2018}, we find that the effect of the radial electric field can be large even when the amplitude of the potential flux-surface variation is small relative to the temperature.
In addition, we find that classical transport can dominate over the neoclassical transport for collisional impurities in certain stellarator geometries. 

The remainder of this paper is organised as follows:
in \autoref{sec:imp}, we present the equations describing the impurities, and relate the friction force acting on the impurities to their flux-surface density variations and the resulting radial flux.
In \autoref{sec:ff}, we introduce the ion-impurity collision operator and obtain an explicit expression for the ion-impurity friction force. In \autoref{sec:smallDelta}, we consider simplifying limits of the equations presented in the previous sections, and derive expressions for transport coefficients in those limits. \autoref{sec:class} treats the classical transport, and shows why it is important in Wendelstein 7-X.
Finally, in \autoref{sec:w7xtest}, we apply our results to study a test-case based on a Wendelstein 7-X vacuum field. 

\section{Impurity equations}
\label{sec:imp}
In this section, we present equations to model the impurities, starting from momentum balance and ending with expressions for calculating the flux along the magnetic field and across the flux-surface.

The impurities are assumed to be collisional enough to be in the Pfirsh-Schl\"{u}ter regime and thus have a Maxwellian velocity distribution, with the density not necessarily constant on flux-surfaces. For such a species in steady-state, the momentum equation is
\begin{equation}
\nabla p_z = Z e n_z \vec{E} + Z e \vec{\Gamma}_z \times \vec{B} + \vec{R}_z  \label{eq:momentum}
\end{equation}
where the $z$ species subscript refers to the impurities, $Z$ is the impurity charge-number, $e$ the proton charge, $p_z$ the impurity pressure, $n_z$ the impurity density, $\vec{\Gamma}_z$ the impurity particle flux, $\vec{B}$ the magnetic field, $\vec{E}$ the electric field, and
$\vec{R}_z$ is the friction force acting on the impurities. By projecting \eqref{eq:momentum} onto the magnetic field direction $\vec{b} = \vec{B}/B$, with $B = |\vec{B}|$, we obtain
\begin{equation}
\nabla_\| p_z  =  Z e n_z E_\| + R_{z\|}. \label{eq:parallel}
\end{equation}
From \eqref{eq:parallel}, we see that pressure (and thus density) variation along the field-line is set up by forces associated with the parallel electric field and friction -- both of which increase with the impurity charge number. 
The friction force can be calculated using kinetic information of all other species, as
\begin{equation}
\vec{R}_{z} = m_z \sum_a \int \!\d^3v\,  \vec{v} C[f_z,f_a], 
\end{equation}
where $f_a$ is the distribution function of a species ``$a$''. %, and the sum goes over all species. 
We will restrict ourselves to the case where only collisions between a bulk ion species $i$ and the impurities matter:  the electron contribution to the friction force can be neglected as small in the electron-ion mass ratio.

In order to simplify the kinetic calculations required to determine $f_i$, we will assume that $Z \gg 1$, so that the effects that lead to pressure variation on the flux-surface can be significant for the impurities while being small for the bulk ions. In the $Z \gg 1$ limit, the ions and impurities will have undergone temperature equilibration if \citep{helanderBifurcated1998}
\begin{equation}
\frac{\rho_* \hat{\nu}_{ii}}{Z} \ll 1, \label{eq:Teq}
\end{equation}
where $\rho_* = \rho_i/L$, with $L$ the profile length-scale and $\rho_i=v_{Ti}m_i/eB$ the ion thermal gyroradius, with $m_i$ the ion mass and $v_{Ti} = \sqrt{2T_i/m_i}$ the ion thermal speed; $\hat{\nu}_{ii}=n_i e^4 \ln\Lambda\, L_\|/(T_{i}^{2} \epsilon_0^2 12 \pi^{3/2})$ is the ion collisionality, where $n_i$ is the bulk-ion density, the bulk ions are assumed to have $Z=1$; $L_\|$ is the length-scale of $\Phi$-variations parallel to $\vec{B}$, where we assume that the inductive electric field is small, so that $\vec{E}=-\nabla \Phi$; $\epsilon_0$ is the vacuum permittivity and $\ln\Lambda$ the Coulomb-logarithm.
\autoref{eq:Teq} is practically always satisfied in a magnetized plasma, so we will assume that $T_z = T_i$ is a flux-function, and \eqref{eq:parallel} thus becomes an equation for the flux-surface variation of $n_z$.

Furthermore, if $\Delta \equiv Z^2 \rho_* \hat{\nu}_{ii}  \ll 1$, as in the conventional drift-kinetic ordering,  the friction force in \eqref{eq:parallel} becomes smaller than the other terms \citep{helanderBifurcated1998}. To zeroth order in $\Delta$, the density in \eqref{eq:parallel} is then given by a Boltzmann response to $\Phi$
\begin{equation}
n_z = N_z \e^{-Ze\Phi/T_z}, \qquad (\Delta \ll 1) \label{eq:boltzmann}
\end{equation}
where $N_z$ is a flux-function. If the density variation of all species is given by \eqref{eq:boltzmann}, quasi-neutrality forces the density and potential to also be flux-functions. For significant density variation to arise on a flux-surface, the behaviour of at least one species must thus deviate from \eqref{eq:boltzmann}; several different mechanisms have been considered in the literature:

In \nouncite{helanderBifurcated1998}, $\Delta = \ordo{1}$, so the impurities themselves set up their own flux-surface variation to balance the flux-surface variation of the friction force and electric field. This was generalized in \nouncite{fulop1999} to include centrifugal forces.

Additionally, heating can introduce a fast particle population, which may not have a density variation according to \eqref{eq:boltzmann} and thus leads to an electric field tangential to the flux surface, which the impurity density in \eqref{eq:parallel} responds to. Such effects were considered in \nouncite{kazakov2012} and \nouncite{angioni2014}, and are often more important than the variations set up by the impurities themselves. %\todo{Cite what for smallness of intrinsic effect?}

Furthermore, in stellarators, helically trapped particles drifting due to the radial electric field can cause flux-surface density asymmetries that, in turn, causes an electric field \citep{garcia2017}. This electric field then affects the impurities. This mechanism has been investigated numerically in \nouncite{garcia2017}, and was found to significantly affect the transport in the Large Helical Device (LHD) and TJ-II stellarators, but does not appear to have a major effect in Wendelstein 7-X (W7-X) due to the neoclassical optimization reducing the radial extent of such helically trapped orbits.

Out of these mechanisms, the latter two are expected to be more important. For the sake of generality, we will however allow $\Phi$ and $\Delta$ to be arbitrary, as long as the tangential variation in $\Phi$ (which we denote by $\tilde{\Phi}$), is of magnitude  $e\tilde{\Phi}/T_i \sim Z^{-1}$ so that its effect can be neglected for the bulk ions. 
In \autoref{sec:smallDelta} and later sections, we will consider the case when $\Delta \ll 1$, with $\tilde{\Phi}$ centered around extrema of $B$.

\subsection{Radial impurity flux}
Regardless of the mechanisms that determine the spatial variation of $n_z$, we can calculate the perpendicular flux of the Maxwellian impurities by applying $\vec{B} \times$ to \eqref{eq:momentum}, resulting in
\begin{equation}
\begin{aligned}
\vec{B} \times \nabla p_z = \vec{B} \times Z e n_z \vec{E} + Z e B^2\vec{\Gamma}_{z,\perp}  + \vec{B} \times \vec{R}_z.  \label{eq:Bxmomentum}
\end{aligned}
\end{equation}
This expression contains the flux in both the diamagnetic and radial directions. The flux-surface averaged radial flux becomes
\begin{equation}
\begin{aligned}
&Z e \lang \vec{\Gamma}_z \cdot \nabla \psi \rang =  \lang \frac{\vec{B} \times \nabla \psi}{B^2} \cdot\vec{R}_z \rang  \\
& +  Z e \lang n_z \frac{\vec{B} \times \nabla \psi}{B^2} \cdot \vec{E} \rang -\lang \frac{\vec{B} \times \nabla \psi}{B^2} \cdot \nabla p_z \rang, 
\end{aligned} \label{eq:radial}
\end{equation}
where $\psi$ is an arbitrary flux-surface label and $\lang \cdot \rang$ denotes the flux-surface average. Here, the first term on the right is the classical flux, and the second one is the radial flux due to the $E \times B$-drift. As there is no radial current in steady-state (i.e.\ $\nabla \times \vec{B} \cdot \nabla \psi = 0$), we have
\begin{equation}
  \lang \vec{B} \times \nabla \psi \cdot \nabla X \rang = 0
\end{equation}
for any single-valued function $X$. The last term of \eqref{eq:radial} can thus be rewritten as $\lang p_z \vec{B} \times \nabla \psi \cdot \nabla B^{-2} \rang$, which is the radial flux due to the magnetic drift of a Maxwellian species. The two latter terms in \eqref{eq:radial} thus correspond to the neoclassical flux, and will be denoted by $\lang \vec{\Gamma}_z \cdot \nabla \psi \rang^{\text{NC}}$.

%From this point, we will assume that the inductive electric field is small, so that $\vec{E} = -\nabla \Phi$.
Following \nouncite{calvo2018}, we obtain a flux-friction relation by introducing the function  $w$\footnote{Note that \citet{calvo2018} defines $U_1$ instead of $w$; they are related through $U_1 = w n_z/N_z$.}, defined through the magnetic differential equation
\begin{align}
  %\vec{B} \cdot \nabla u &= - \vec{B} \times \nabla \psi \cdot \nabla B^{-2} \\
  \vec{B} \cdot \nabla (n_z w) &= - \vec{B} \times \nabla \psi \cdot \nabla (n_z B^{-2}), \label{eq:w}
\end{align}
so that
\begin{equation}
\begin{aligned}
  Ze\lang \vec{\Gamma}_z \cdot \nabla \psi \rang^{\text{NC}} &=\lang \vec{B} \cdot [Ze n_z w \nabla \Phi + w T_i n_z \nabla \ln n_z]\rang \\
  &= \lang B  w R_{z\|}\rang
\end{aligned} \label{eq:radial2}
\end{equation}
where we have used parallel force-balance to relate the gradients to the friction force $R_{z\|}$.
An expression for $R_{z\|}$ is presented in \autoref{sec:ff}. To calculate the friction force, we must however know the parallel impurity flux, which is the subject of the next section.

\subsection{Parallel impurity flux}
\label{sec:pif}
From \eqref{eq:Bxmomentum}, we get the impurity flux in the $\vec{B} \times \nabla \psi$-direction (denoted with a $\wedge$ subscript) as
\begin{equation}
\begin{aligned}
\vec{\Gamma}_{z\wedge}= \frac{\vec{B} \times \nabla \psi}{ZeB^2} \left( Z e n_z \frac{\p \Phi}{\p \psi} +\frac{\p p_z}{\p \psi}   -  \frac{\vec{R}_{z} \cdot\nabla \psi}{|\nabla \psi|^2} \right).  \label{eq:dia}
 \end{aligned}
\end{equation}
In a confined plasma, the radial fluxes and thus the radial friction will be small, so we can neglect $\vec{R}_{z} \cdot \nabla \psi$ in \eqref{eq:dia} and neglect the radial flux in the impurity continuity equation $\nabla \cdot \vec{\Gamma}_z = 0$. The parallel impurity flux $\Gamma_{z,\|}$ thus satisfies
 \begin{equation}
 \vec{B} \cdot \nabla (\Gamma_{z\|}B^{-1}) = - \nabla \cdot \vec{\Gamma}_{z\wedge}. \label{eq:Gpar1}
\end{equation}
%In \eqref{eq:dia}, the second term will be small when $Z \gg 1$, unless $p_z$ varies on scales $Z$ times shorter than $\Phi$. Such large $n_z$ variation are indeed predicted if the impurity profiles are evolved towards equilibrium ($\lang \vec{\Gamma}_z \cdot \nabla \psi \rang = 0$, see e.g.\ \citet{helander2005} or \citet{calvo2018}).  We will nevertheless still assume that the second term is small, which should be a valid assumption before the impurities have reached their long-time equilibrium. Depending on the specifics of the discharge, this may not be such a large limitation, as the accumulation of impurities may terminate the discharge before this equilibrium is reached.

In the $Z \gg 1$ limit, \eqref{eq:Gpar1} thus becomes (recalling $e\tilde{\Phi}/T \sim Z^{-1}$)
 \begin{equation}
 \vec{B} \cdot \nabla (\Gamma_{z\|}B^{-1}) = - \frac{\d \lang \Phi \rang}{\d \psi}\vec{B} \times \nabla \psi \cdot \nabla \left( \frac{n_z}{B^2}\right) - \frac{T_i}{Ze}\vec{B} \times \nabla \psi \cdot \nabla \left( \frac{1}{B^2} \frac{\p n_z}{\p \psi}\right),
\end{equation}
where we have retained $\frac{\p n_z}{\p \psi}$ to account for the fact that steady-state impurity density profiles tend to be $Z$ times larger than those of the bulk ion, i.e.\ $\p_\psi n_z/n_z \sim Z \d_\psi n_i/n_i \sim Z \d_\psi T_i/T_i$ \citep{helander2005, calvo2018}.

In the $\Delta \ll 1$ limit, we can use \eqref{eq:boltzmann} to obtain an explicit expression for $\p_\psi n_z$, resulting in

\begin{equation}
\Gamma_{z\|} \equiv n_z V_{z\|}= w n_z \left(\frac{\d \lang \Phi \rang}{\d \psi} + \frac{T_i}{Ze N_z} \frac{\d N_z}{\d \psi}\right)B + BK_z, \qquad (\Delta \ll 1)\label{eq:Gpar2}
\end{equation}
where $K_z(\psi)$ is an integration constant, and we have dropped $\ordo{Z^{-1}}$ terms. %The results can be simplified in various limits:

\section{Parallel friction force}
\label{sec:ff}
With the parallel impurity flux from the previous section, we now have everything needed to calculate the ion-impurity parallel friction.

As the collisions with electrons can be neglected, the friction force on the impurities can be expressed as
\begin{equation}
\vec{R}_{z} \approx \vec{R}_{zi} = -\vec{R}_{iz} = - \int \d^3 v m_i \vec{v} C_{iz},
\end{equation}
where $\vec{R}_{ab}$ denotes the friction force on species $a$ by species $b$, and $C_{iz}$ is the ion-impurity collision operator. Since $m_z \gg m_i$ for high-$Z$ impurities, we can use a mass-ratio expanded ion-impurity collision operator
\begin{equation}
C_{iz} = \nu_{iz}^D(v) \left(\mathcal{L}(f_{i1}) + \frac{m_i \vec{v} \cdot \vec{V}_{z}}{T_i} f_{i0}\right), \label{eq:ciz}
\end{equation}
where $\vec{V}_{z} = \vec{\Gamma}_z/n_z$ is the flow of the impurities, $\mathcal{L}$ is the Lorentz operator \citep{helander2005}, $f_{i1}$ the order $\rho_*$ part of the ion distribution function, and the collision frequency $\nu_{iz}^D$ is
\begin{equation}
\nu_{iz}^D= \frac{n_z Z^2 e^4 \ln \Lambda}{4\pi m_i^2 \epsilon_0^2 v^3}.
\end{equation}
The lowest order ion distribution function $f_{i0}$ is taken to be a stationary Maxwell-Boltzmann distribution, and to calculate the parallel friction force we only need the gyrophase-independent part of $f_{i1}$ (which we denote by $F_{i1}$). This function is given by the ion drift-kinetic equation
\begin{equation}
\begin{aligned}
v_\| \nabla_\| F_{i1} + \vec{v}_d \cdot \nabla f_{i0}  = C_i,
\end{aligned}\label{eq:dke}
\end{equation}
where gradients are taken with $\E = mv^2/2 + e\Phi$ and $\mu = m_i v_\perp^2/(2B)$ fixed -- although we will later make use of the fact that the potential energy is approximately constant over an ion orbit and use the approximate invariants $v$ and $\lambda = v_\perp^2/(v^2B)$ as velocity coordinates.

The collision operator is approximately given by collisions with bulk ions and impurities, $C_i \approx C_{iz} + C_{ii}$, and we use a model operator for ion-ion collisions 
\begin{equation}
C_{ii} =  \nu_{ii}^D(v) \left(\mathcal{L}(F_{i1}) + \frac{m_i \vec{v} \cdot \vec{U}}{T_i} f_{i0}\right), \label{eq:cii}
\end{equation}
where $\vec{U}$ is determined by momentum conservation and the collision frequency is 
\begin{equation}
\nu_{ii}^D = \frac{n_i e^4 \ln \Lambda}{4\pi m_i^2 \epsilon_0^2  v^3} \left(\erf{(v/v_{Ti})} - G(v/v_{Ti})\right),
\end{equation}
where $\erf$ is the error function and $G$ the Chandrasekhar function \citep{helander2005}.

We will assume that $C_{i}$ is smaller than the other terms in \eqref{eq:dke} and expand $F_{i1} =F_{i1(-1)}  + F_{i1(0)} + F_{i1(1)} + \dots $ in collisionality, so that
\begin{align}
  &v_\| \nabla_\| F_{i1(-1)}  = 0 \label{eq:fi1m1}\\
  &v_\| \nabla_\| F_{i1(0)} + \vec{v}_d \cdot \nabla f_{i0}   = C_i[F_{i1(-1)}] \label{eq:fi10}\\
  &v_\| \nabla_\| F_{i1(1)}  = C_i[F_{i1(0)}] \label{eq:fi11}.
\end{align}
We solve \eqref{eq:fi1m1}, \eqref{eq:fi10} and \eqref{eq:fi11} as in \nouncite{newton2017}, except that we allow $n_z$ to vary on the flux-surface, which makes the expressions less compact; the details are thus relegated to appendices \ref{sec:sidke}--\ref{sec:solvcond}.

The parallel friction force becomes
\begin{equation}
  \begin{aligned}
    R_{iz\|} =  & \frac{n_i m_i}{\tau_{iz}} \left(V_{z\|} - \frac{T_i}{e} \left[ A_{i1} - \frac{3}{2} A_{i2}\right] Bu - BP(\psi) \right), \label{eq:pFF}
  \end{aligned}
\end{equation}

% \begin{equation}
%   \begin{aligned}
% R_{iz,\|} \frac{\tau_{iz}}{n_i m_i} =&  \left[\w - \frac{\lang \w B^2 \rang}{\lang B^2 \rang}\right] B \frac{\d \lang \Phi \rang}{\d \psi}  
% + \left[\frac{1}{n_z} - \frac{\lang \frac{B^2}{n_z} \rang}{\lang B^2 \rang}\right] BK_z 
% \\&- \left[u -\frac{\lang u B^2 \rang}{\lang B^2 \rang}\right] B \frac{T_i}{e}  \left[A_{i1} - \frac{3}{2} A_{i2}\right],
% \end{aligned}
% \label{eq:RizKz}
% \end{equation}
where $A_{i1} = \frac{\d \ln p_i}{\d \psi} + \frac{e}{T_i} \frac{\d \lang \Phi \rang}{\d \psi}$ and $A_{2i}  = \frac{\d \ln T_i}{\d \psi}$ are the ion thermodynamic forces; $\tau_{iz}^{-1} = Z^2 n_z e^4 \ln \Lambda/(3\pi^{3/2} m_i^2 \epsilon_0^2 v_{Ti}^3)$; $P(\psi)$ is a flux-surface constant defined in \eqref{eq:P1}; $V_{z\|}$ is obtained from \eqref{eq:Gpar1} combined with the solvability condition to \eqref{eq:parallel}, as described in \autoref{sec:solvcond}; $u$ satisfies the magnetic equation
\begin{align}
\vec{B} \cdot \nabla u &= - \vec{B} \times \nabla \psi \cdot \nabla B^{-2}, \label{eq:u}
\end{align}
with $u=0$ at the maximum of $B$.

\autoref{eq:pFF} can be used to solve for $n_z$ from the parallel momentum equation \eqref{eq:parallel}, given a mechanism to set $\tilde{\Phi}$. We will not attempt such a daunting task at this time, and instead consider the $\Delta \ll 1$ limit in the following section.
In this limit, the form of $V_{z\|}$ is known from \eqref{eq:Gpar2}, so the parallel friction becomes
\begin{equation}
  \begin{aligned}
R_{iz,\|} \frac{n_z \tau_{iz}}{n_i m_i} =&  \left[n_z \w B - n_z B \frac{\lang \w B^2 \rang}{\lang B^2 \rang}\right]  \frac{T_i}{e} \left(\frac{e}{T_i}\frac{\d \lang \Phi \rang}{\d \psi} + \frac{1}{ZN_z} \frac{\d N_z}{\d \psi}\right)  
+ \left[B - n_z B \frac{\lang \frac{B^2}{n_z} \rang}{\lang B^2 \rang}\right] K_z 
\\&- \left[n_z u B-n_z B\frac{\lang u B^2 \rang}{\lang B^2 \rang}\right] \frac{T_i}{e}  \left[A_{i1} - \frac{3}{2} A_{i2}\right],
\end{aligned}
\label{eq:RizKz}
\end{equation}
where $K_z$ is determined by the solvability condition, and is given by
\begin{align}
 K_z(\psi)
= & 
- \lang\frac{B^2}{n_z} (1 - c_4 \alpha) \rang^{-1}  \lang (1 - c_4\alpha) \w B^2 \rang  \frac{T_i}{e} \left( \frac{e}{T_i}\frac{\d \lang \Phi \rang}{\d \psi}  + \frac{1}{ZN_z} \frac{\d N_z}{\d \psi}\right)\nonumber\\
&+\lang\frac{B^2}{n_z} (1 - c_4 \alpha) \rang^{-1}\left(c_2  + \lang u B^2 \rang \left[c_1 + 1\right]\right) \frac{T_i}{e} A_{1i}
 \label{eq:Kz1} \\ &+ \lang\frac{B^2}{n_z} (1 - c_4 \alpha) \rang^{-1}\left( c_3 - \frac{5}{2} c_2 
- \lang u B^2 \rang\left[c_1 \eta + \frac{3}{2} \right]
\right)\frac{T_i}{e}A_{2i}, \nonumber
\end{align}
where $\alpha = Z^2 n_z/n_i$; $\eta \approx 1.17$; the $c_i$ are flux-surface constants which depend on the magnetic geometry and the impurity density variations on the flux-surface, and are defined in equations \eqref{eq:c1}--\eqref{eq:c4}.

% \subsection{Trace impurity limit}
% In the trace impurity limit, $\alpha \ll 1$, all the non-standard flux-surface constants -- $a_i$, $b_j$ and $c_k$ -- reduce to known functions that only depend on the magnetic geometry, see \eqref{eq:tra1}--\eqref{eq:trb4} in \autoref{sec:abtrace}. The resulting friction force is
% \begin{align}
% R_{iz,\|} &\frac{\tau_{iz}}{n_i m_i} =  \left(\w  - \frac{\lang \w B^2 \rang }{n_z \lang\frac{B^2}{n_z} \rang} \right) B \frac{\d \lang \Phi \rang}{\d \psi}   \label{eq:Fiztrace}\\
% &+ \left(\frac{\lang u B^2 \rang}{\lang B^2 \rang}-u + \left(\frac{1}{n_z \lang\frac{B^2}{n_z} \rang} - \frac{1}{\lang B^2 \rang}\right)  \left[f_s + \lang uB^2 \rang \right] \left[\frac{f_c}{1 - f_c} + 1 \right] \right) B \frac{T_i}{e} A_{i1} \nonumber\\
% &- \left(\frac{3}{2}\frac{\lang u B^2 \rang}{\lang B^2 \rang}-\frac{3}{2}u + \left(\frac{1}{n_z \lang\frac{B^2}{n_z} \rang} - \frac{1}{\lang B^2 \rang}\right)  \left[f_s + \lang uB^2 \rang \right] \left[\frac{\eta f_c }{1 - f_c} + \frac{3}{2} \right] \right)   B \frac{T_i}{e} A_{i2}, \nonumber
% \end{align}
% where $f_c$ and $f_s$ are defined in \eqref{eq:fc} and \eqref{eq:fs}.
% \autoref{eq:Fiztrace} can be used to solve for $n_z$ from the parallel momentum equation \eqref{eq:parallel}, given a mechanism to set $\tilde{\Phi}$. We will not attempt such a daunting task at this time, and instead consider simplifying limits in the following section.

\section{Impurities in the $\Delta \ll 1$ limit}
\label{sec:smallDelta}
In the $\Delta \ll 1$ limit, with $n_z = n_{z0} + n_{z1} + \dots$ and $\tilde{\Phi} = \tilde{\Phi}_0 + \tilde{\Phi}_1 + \dots$,
the zeroth-order parallel momentum equation becomes
\begin{equation}
T_z \nabla_\| n_{z0}  =  -Z e n_{z0} \nabla_\| \tilde{\Phi}_{0},
\end{equation}
so the zeroth order impurity density is given by a Boltzmann response to $\tilde{\Phi}_0$
\begin{equation}
n_{z0}  =  N_z(\psi) \e^{-Ze\tilde{\Phi}_{0}/T_z}, \label{eq:boltzmann2}
\end{equation}
where $N_z$, sometimes referred to as the \emph{pseudo-density}, is a flux-function.
Here, we assume that $\tilde{\Phi}_0$ is known and set by a mechanism unrelated to flux-surface variation in $n_z$. This is appropriate, since we know that $n_{z0}$ cannot give rise to a non-zero $\tilde{\Phi}$, so that $n_{z}$ gives no contribution to $\tilde{\Phi}$ to this order; recall the discussion below \eqref{eq:boltzmann}.

%This corresponds to $\tilde{\Phi}$ being set by a mechanism unrelated to $n_z$.  %-- such as fast particles -- and receives a negligible contribution from $n_z$ This is appropriate in the trace limit, as $Z^2 n_z \ll n_i$ so that $Zn_z$ can be neglected in the quasi-neutrality relation, which would otherwise relate $\tilde{\Phi}$ and $n_z$. \autoref{eq:boltzmann2} is also appropriate in the non-trace limit, but $\tilde{\Phi}$ can be replaced by $\tilde{\Phi}_{0}$, which excludes effects due to higher order corrections to $n_z$.

If \eqref{eq:boltzmann2} is used to write $\tilde{\Phi}_0$ in terms of $n_{z0}$, the first-order parallel momentum equation becomes
\begin{equation}
T_z n_{z0} \nabla_\| \left( \frac{n_{z1}}{n_{z0}}  + \frac{Ze}{T_z}\tilde{\Phi}_1\right)  = R_{z\|}[n_{z0}], \label{eq:Delta1}
\end{equation}
which has the solvability condition $\lang n_{z0}^{-1} B  R_{z\|}[n_{z0}] \rang =0$. This is the same solvability condition as that of the exact equation \eqref{eq:parallel}, except with $n_z \to n_{z0}$, which implies that the flux to order $\Delta^1$ can be consistently calculated from \eqref{eq:radial2} with $n_{z0}$. %Thus, there is no need to calculate $n_{z1}$.

We thus have
\begin{equation}
 \lang \vec{\Gamma}_z \cdot \nabla \psi \rang^{\text{NC}} = \frac{1}{Ze}\lang  w_0 B R_{z\|}[n_{z0}] \rang, \label{eq:zfluxf}
 \end{equation}
 where $w_0$ is given by \eqref{eq:w} but with $n_z \to n_{z0}$, and $-R_{z\|}$ is given by \eqref{eq:RizKz}. The resulting flux can be written
\begin{equation}
\frac{\lang \vec{\Gamma}_z \cdot \nabla \psi \rang^{\text{NC}}}{\lang n_{z0} \rang } = D_\Phi^{\text{NC}} \frac{e}{T_i} \frac{\d \lang \Phi \rang}{\d \psi} - \frac{1}{Z}D_{N_z}^{\text{NC}} \frac{\d \ln N_z}{\d \psi} - D_{n_i}^{\text{NC}}  \frac{\d \ln n_i}{\d \psi} - D_{T_i}^{\text{NC}}  \frac{\d \ln T_i}{\d \psi}, \label{eq:fluxDnc}
\end{equation}
where
\begin{align}
  D_\Phi^{\text{NC}} = &
  \frac{m_i n_i T_i}{Ze^2 \lang n_{z0} \rang n_{z0} \tau_{iz0} }\left[\vphantom{\frac{\frac{\lang n_{z0} w_0 B^2 \rang}{\lang B^2\rang} \lang \frac{B^2}{n_{z0}}\rang-\lang w_0B^2 \rang }{\lang \frac{B^2}{n_{z0}} (1 - c_4 \alpha)\rang}}
                                                                        \lang n_{z0} w_0 (u-w_0)B^2\rang - \frac{\lang n_{z0} w_0 B^2 \rang}{\lang B^2 \rang} \lang  (u-w_0) B^2 \rang 
  \right. \label{eq:DPhi}
  \\&+\left.
    \frac{\frac{\lang n_{z0} w_0 B^2 \rang}{\lang B^2\rang} \lang \frac{B^2}{n_{z0}}\rang-\lang w_0B^2 \rang }{\lang \frac{B^2}{n_{z0}} (1 - c_4 \alpha)\rang}
      \left(c_2 + \lang (u-w_0)B^2\rang + c_1 \lang uB^2 \rang + c_4 \lang \alpha w_0 B^2  \rang\right) \right]
      \nonumber\displaybreak[0]\\
  D_{N_z}^{\text{NC}} =& \frac{m_i n_i T_i}{Z e^2 \lang n_{z0} \rang n_{z0} \tau_{iz0} }\left[\vphantom{\frac{\frac{\lang n_{z0} w_0 B^2 \rang}{\lang B^2\rang} \lang \frac{B^2}{n_{z0}}\rang-\lang w_0B^2 \rang }{\lang \frac{B^2}{n_{z0}} (1 - c_4 \alpha)\rang}}
                         \lang n_{z0} w_0^2 B^2 \rang  - \lang n_{z0} w_0B^2 \rang\frac{\lang w_0 B^2 \rang}{\lang B^2 \rang}  \right. \label{eq:DNz}\\
&+\left. \frac{\frac{\lang n_{z0} w_0 B^2 \rang}{\lang B^2\rang} \lang \frac{B^2}{n_{z0}}\rang-\lang w_0B^2 \rang }{\lang \frac{B^2}{n_{z0}} (1 - c_4 \alpha)\rang}  \lang (1 - c_4 \alpha)w_0B^2\rang \right]\nonumber\displaybreak[0]\\
  D_{n_i}^{\text{NC}} = &
                          -\frac{m_i n_i T_i}{Ze^2 \lang n_{z0} \rang n_{z0} \tau_{iz0} Ze}\left[\vphantom{\frac{\frac{\lang n_{z0} w_0 B^2 \rang}{\lang B^2\rang} \lang \frac{B^2}{n_{z0}}\rang-\lang w_0B^2 \rang }{\lang \frac{B^2}{n_{z0}} (1 - c_4 \alpha)\rang}}
                          \lang n_{z0} w_0 u B^2\rang - \lang n_{z0} w_0 B^2\rang \frac{\lang u B^2 \rang}{\lang B^2 \rang}
       \right. \label{eq:Dni}
  \\&\qquad+\left.
      \frac{\frac{\lang n_{z0} w_0 B^2 \rang}{\lang B^2\rang} \lang \frac{B^2}{n_{z0}}\rang-\lang w_0B^2 \rang }{\lang \frac{B^2}{n_{z0}} (1 - c_4 \alpha)\rang} \left(c_2 + \lang uB^2 \rang [c_1 +1]\right)
   \right]\nonumber\displaybreak[0]
\\
  D_{T_i}^{\text{NC}} = &
                          \frac{m_i n_i T_i }{Ze^2\lang n_{z0} \rang n_{z0} \tau_{iz0} }\left[
                          \vphantom{\frac{\frac{\lang n_{z0} w_0 B^2 \rang}{\lang B^2\rang} \lang \frac{B^2}{n_{z0}}\rang-\lang w_0B^2 \rang }{\lang \frac{B^2}{n_{z0}} (1 - c_4 \alpha)\rang}}
\frac{1}{2}\left( \lang n_{z0} w_0 u B^2\rang - \lang n_{z0} w_0 B^2\rang \frac{\lang u B^2 \rang}{\lang B^2 \rang}\right)
                          \right.\label{eq:DTi}
  \\&\qquad-\left.
      \frac{\frac{\lang n_{z0} w_0 B^2 \rang}{\lang B^2\rang} \lang \frac{B^2}{n_{z0}}\rang-\lang w_0B^2 \rang }{\lang \frac{B^2}{n_{z0}} (1 - c_4 \alpha)\rang}
  \left(c_3 - \frac{3}{2} c_2 - \lang uB^2 \rang \left[c_1(\eta - 1) + \frac{1}{2}\right]\right)\right],  \nonumber
\end{align}
with $\tau_{iz0}$ given by the expression for $\tau_{iz}$, but with $n_z \to n_{z0}$. From \eqref{eq:DPhi}, we see that the flux due to the radial electric field is generally non-zero, but that it vanishes when $n_{z0}$ is constant on the surface. The non-zero $D_\Phi^{\text{NC}}$ can in fact dominate the other neoclassical transport coefficients, as will be seen in \autoref{sec:w7xtest}.

%% From \eqref{eq:Delta1} and \eqref{eq:pFF}, $n_{z1}$ can now be obtained for a given $\tilde{\Phi}_0$ or $n_{z0}$, and an expression for $\tilde{\Phi}_1$ in terms of $n_{z1}$. However, neither $n_{z1}$ or $\tilde{\Phi}_1$ is actually required to calculate the flux, as \eqref{eq:Delta1} and \eqref{eq:radial2} directly gives that
%% \begin{equation}
%% \lang \vec{\Gamma}_z \cdot \nabla \psi \rang^{\text{NC}} = \frac{1}{Ze}\lang  w_0 B R_{z\|}[n_{z0}] \rang, \label{eq:zfluxf}
%% \end{equation}
%% where $w_0$ is given by \eqref{eq:w} but with $n_z \to n_{z0}$.

% One remaining difficulty is evaluating $\omega$. The problem is that $N_z(\psi)$ is unknown, which makes it impossible to evaluate $\frac{\p n_z}{\p \psi}$ required to calculate $\w$. In light of \eqref{eq:boltzmann}, we can see that the assumption of weak diamagnetic flow in \autoref{sec:pif} corresponds to a sharply varying $N_z(\psi)$ canceling the sharp variation due to $Ze \Phi$ in the exponential. In contrast, when $N_z$ does not have radial gradients of order $Z$, the diamagnetic and $E \times B$ fluxes cancel in the $\vec{B} \times \nabla \psi$-direction, and $\w =0$. We will use the former assumption, and thus $\w = w_0$.

\subsection{Trace limit}
The transport coefficients in \eqref{eq:DPhi} -- \eqref{eq:DTi} simplify somewhat in the trace limit, where all the $c_i$ reduce the expressions in terms of standard functions of geometry.
Using \eqref{eq:tc1} -- \eqref{eq:tc4}, we get that
  % \begin{align}
%     &\frac{\tau_{iz0} n_{z0}}{n_i m_i} \lang w_0 B R_{z\|}[n_{z0}] \rang = 
%     \left(\frac{\lang w_0B^2 \rang^2}{\lang \frac{B^2}{n_{z0}} \rang}  - \lang n_{z0} w_0^2 B^2 \rang  \right)  \frac{T_i}{e} \left(\frac{\d \lang \Phi \rang}{\d \psi} + \frac{1}{ZN_z}\frac{\d N_z}{\d \psi}\right)  \label{eq:fluxnc2}\\
%     &+\left(\lang n_{z0} w_0 uB^2 \rang -\frac{\lang u B^2 \rang}{\lang B^2 \rang}\lang n_{z0} w_0B^2 \rang \right)
% \frac{T_i}{e}
%   \left(A_{i1} - \frac{3}{2} A_{i2}\right) \nonumber\\
%   &+ \left(\frac{ \lang n_{z0} w_0B^2 \rang}{\lang B^2 \rang} - \frac{\lang  w_0B^2\rang}{\lang\frac{B^2}{n_{z0}} \rang}\right) \left(f_s + \lang uB^2 \rang \right) \frac{T_i}{e} \left( \frac{f_c}{1 - f_c}   \left[A_{1i} - \eta  A_{2i}\right] +  \left[A_{i1} - \frac{3}{2} A_{i2}\right]\right), \nonumber
% \end{align}
% which we can decompose into parts driven by density, temperature and potential gradients
% \begin{equation}
% \frac{\lang \vec{\Gamma}_z \cdot \nabla \psi \rang^{\text{NC}}}{\lang n_{z0} \rang } = D_\Phi^{\text{NC}} \frac{e}{T_i} \frac{\d \lang \Phi \rang}{\d \psi} - D_{n_i}^{\text{NC}}  \frac{\d \ln n_i}{\d \psi} - D_{T_i}^{\text{NC}}  \frac{\d \ln T_i}{\d \psi} - D_{N_z}^{\text{NC}} \frac{\d \ln N_z}{\d \psi}, 
% \end{equation}
% where
\begin{align}
  D_\Phi^{\text{NC}} = 
    \frac{m_i n_i T_i}{Ze^2 \lang n_{z0} \rang n_{z0} \tau_{iz0}}&\left( \lang n_{z0} w_0 (u-w_0)B^2 \rang  +\frac{\lang w_0B^2 \rang^2}{\lang \frac{B^2}{n_{z0}} \rang}  
   -\frac{\lang u B^2 \rang}{\lang B^2 \rang}\lang n_{z0} w_0B^2 \rang \right. \nonumber
  \\&\qquad+\left.
  \left[\frac{ \lang n_{z0} w_0B^2 \rang}{\lang B^2 \rang} - \frac{\lang  w_0B^2\rang}{\lang\frac{B^2}{n_{z0}} \rang}\right] \frac{\left(f_s + \lang uB^2 \rang \right)}{1-f_c}\right) 
      \label{eq:tDPhi}\displaybreak[0]\\
    D_{N_z}^{\text{NC}} = \frac{ m_i n_i T_i}{Ze^2\lang n_{z0} \rang n_{z0} \tau_{iz0} }&\left(\lang n_{z0} w_0^2 B^2 \rang  - \frac{\lang w_0B^2 \rang^2}{\lang \frac{B^2}{n_{z0}} \rang}  \right)\label{eq:tDNz}\displaybreak[0]\\
  D_{n_i}^{\text{NC}} = 
    -\frac{m_i n_i T_i}{Z e^2 \lang n_{z0} \rang n_{z0} \tau_{iz0}}&\left(
      \lang n_{z0} w_0 uB^2 \rang -\frac{\lang u B^2 \rang}{\lang B^2 \rang}\lang n_{z0} w_0B^2 \rang \right. \label{eq:tDni}
      \\&\qquad+\left.
  \left[\frac{ \lang n_{z0} w_0B^2 \rang}{\lang B^2 \rang} - \frac{\lang  w_0B^2\rang}{\lang\frac{B^2}{n_{z0}} \rang}\right] \frac{\left(f_s + \lang uB^2 \rang \right)}{1-f_c}\right)
\nonumber\displaybreak[0]\\
  D_{T_i}^{\text{NC}} = 
    \frac{1}{2}\frac{m_i n_i T_i}{Ze^2\lang n_{z0} \rang n_{z0} \tau_{iz0}}&\left(
      \lang n_{z0} w_0 uB^2 \rang -\frac{\lang u B^2 \rang}{\lang B^2 \rang}\lang n_{z0} w_0B^2 \rang \right.\label{eq:tDTi}
      \\&\qquad+\left. 
  \left[\frac{ \lang n_{z0} w_0B^2 \rang}{\lang B^2 \rang} - \frac{\lang  w_0B^2\rang}{\lang\frac{B^2}{n_{z0}} \rang}\right] \frac{\left(f_s + \lang uB^2 \rang \right)}{1 - f_c} \left(1 + (2\eta - 3)f_c\right)\right),  \nonumber
\end{align}
which are the expressions we will use in \autoref{sec:w7xtest}.

\section{Classical transport}
\label{sec:class}
Finally, we calculate the classical flux, given by the first term in \eqref{eq:radial}.
Using our mass-ratio expanded collision operator and momentum conservation, the perpendicular friction becomes
\begin{equation}
\vec{R}_{zi\perp} = -\int \d^3 v m_i  \nu_{iz}^D(v) \left(\vec{v}_\perp \mathcal{L}(f_{i1}) + \frac{m_i\vec{v}_\perp\vec{v} \cdot \vec{V}_z}{T_i} f_{i0} \right), \label{eq:Fperp}
\end{equation}
where $\vec{v}_\perp = \vec{v} - v_\|\vec{b}$ with $v_\| = \vec{v} \cdot \vec{b}$.

In \eqref{eq:Fperp}, only the gyrophase-dependent part of $f_{i1}$ contributes to the first term, and only the perpendicular impurity flow contributes to the last.
The gyrophase dependent part of $f_{i1}$ (which we denote $\tilde{f}_{i1}$) is given by \citep{hazeltine1973}
\begin{equation}
\tilde{f}_{i1} = -\vec{\rho}_i \cdot \nabla f_{i0}, \label{eq:ftilde}
\end{equation}
where the gyroradius vector is
\begin{equation}
\vec{\rho}_i = \rho_i (\vec{e}_2 \sin\gamma + \vec{e}_3 \cos \gamma) = \vec{b} \times \vec{v}_\perp/\Omega_i,
\end{equation}
with $\{\vec{b}, \vec{e}_2, \vec{e}_3\}$ an orthonormal set of vectors, and $\Omega_i = eB/m_i$. Thus, we have everything required to calculate the perpendicular friction, which becomes
\begin{equation}
\vec{R}_{zi\perp} =  \frac{m_i n_i}{\tau_{iz}} \frac{T_i}{eB}    \vec{b} \times \nabla \psi \left[A_{1i} - \frac{3}{2} A_{2i}\right] -\frac{n_i m_i}{\tau_{iz}} \vec{V}_{z\perp}.
\end{equation}
Using the same approximations and assumptions as in the neoclassical expressions, we have $\vec{V}_{z\perp} = \frac{T_i}{eB} \left(\frac{e}{T_i} \frac{\d \lang \Phi \rang}{\d \psi} + \frac{1}{ZN_z} \frac{\d N_z}{\d \psi}\right)$, and thus obtain
\begin{equation}
\vec{R}_{zi\perp} =  \frac{m_i n_i}{n_z \tau_{iz}} n_z \frac{\vec{B} \times \nabla \psi}{B^2} \frac{T_i}{e}     \left[\frac{\d \ln n_i}{\d \psi}  - \frac{1}{2}\frac{\d \ln T_i}{\d \psi} - \frac{1}{Z N_z} \frac{\d N_z}{\d \psi} \right],
\end{equation}
resulting in the classical impurity flux
\begin{equation}
  \begin{aligned}
    \lang \vec{\Gamma}_z \cdot \nabla \psi \rang^{\text{C}} \equiv &  \frac{1}{Ze}\lang \frac{\vec{B} \times \nabla \psi}{B^2} \cdot\vec{R}_z \rang \\
    = &  \frac{m_i n_i}{Ze n_z \tau_{iz}} \lang n_z \frac{|\nabla \psi|^2}{B^2}  \rang \frac{T_i}{e}     \left[\frac{\d \ln n_i}{\d \psi}  - \frac{1}{2}\frac{\d \ln T_i}{\d \psi} - \frac{1}{Z N_z} \frac{\d N_z}{\d \psi}\right], \label{eq:fluxc}
\end{aligned}
\end{equation}
or
\begin{equation}
  \begin{aligned}
    \lang \vec{\Gamma}_z \cdot \nabla \psi \rang^{\text{C}} = -\lang n_{z0} \rang \left(\frac{1}{Z}D_{N_z}^{\text{C}} \frac{\d \ln N_z}{\d \psi} + D_{n_i}^{\text{C}} \frac{\d \ln n_i}{\d \psi} +D_{T_i}^{\text{C}} \frac{\d \ln T_i}{\d \psi}\right).
\end{aligned}
\end{equation}

The classical flux is often neglected as smaller than the neoclassical flux. To get a simple estimate of its importance, we take the homogeneous $n_z$ limit of \eqref{eq:fluxDnc} and \eqref{eq:fluxc}, so that the ratio of classical to neoclassical flux depends purely on geometry
\begin{equation}
  \frac{\lang \vec{\Gamma}_z \cdot \nabla \psi \rang^{\text{C}}}{\lang \vec{\Gamma}_z \cdot \nabla \psi \rang^{\text{NC}}} = \frac{\lang \frac{|\nabla \psi|^2}{B^2} \rang \lang B^2 \rang}{\left( \lang u^2B^2 \rang \lang B^2 \rang - \lang u B^2 \rang^2\right)}. \label{eq:ctonc}
\end{equation}
This ratio is indeed small in conventional tokamaks and stellarators (it is $\sim{}\!\!0.1$--$0.6$ in ASDEX Upgrade, and $\sim{}\!\!0.1$--$1$ in LHD), but it is $\sim{}\!\!3$--$3.5$ in a standard W7-X configuration.

W7-X differs from LHD in that it has been optimized to have a low ratio of parallel to perpendicular current. To see how this affects the ratio \eqref{eq:ctonc}, we can express the parallel current in the following way: Charge conservation imposes $\nabla \cdot \vec{j} = 0$, where $\vec{j}$ is the current density. Assuming that the equilibrium magnetic field can be written as $\vec{j} \times \vec{B}  = \nabla p$, where $p$ is the total pressure $p = \sum_a p_a$, which is assumed to be a flux-function to the required order, the parallel current density becomes  $j_\| = uB \frac{\d p}{\d \psi}$. The ratio of parallel and perpendicular current then becomes
\begin{equation}
\frac{j_\|}{|j_\perp|} = \frac{uB}{|\vec{B} \times \nabla \psi/B^2|},
\end{equation}
which can be made small by making $u/|\nabla \psi|$ small, which simultaneously makes \eqref{eq:ctonc} large. The classical flux remains large even when $n_z$ varies on the flux-surface, as we will see in the next section.

\section{Wendelstein 7-X test case}
\label{sec:w7xtest}
\begin{figure}
  \centering
  \begin{subfigure}[b]{0.32\textwidth}
\includegraphics[width=1.0\textwidth]{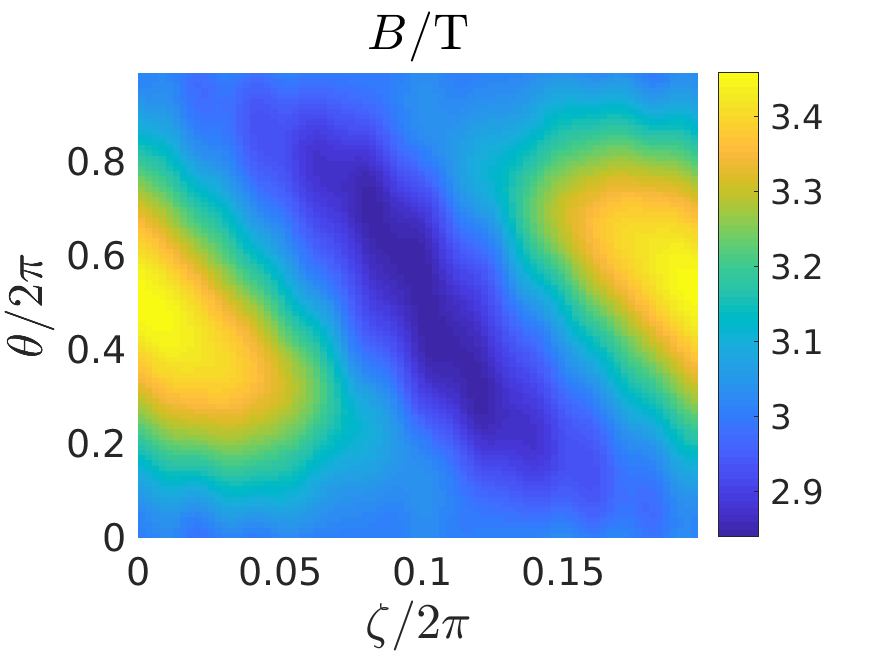}
    %\caption{$B$}
  \end{subfigure}
\begin{subfigure}[b]{0.32\textwidth}
\includegraphics[width=1.0\textwidth]{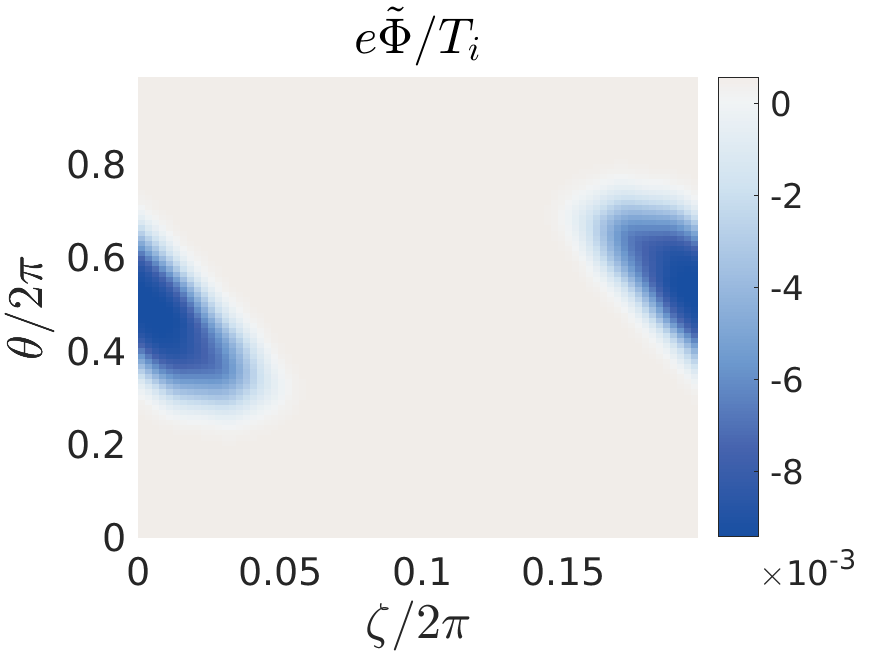}
%\caption{$\tilde{\Phi}$}
  \end{subfigure}
\begin{subfigure}[b]{0.32\textwidth}
\includegraphics[width=1.0\textwidth]{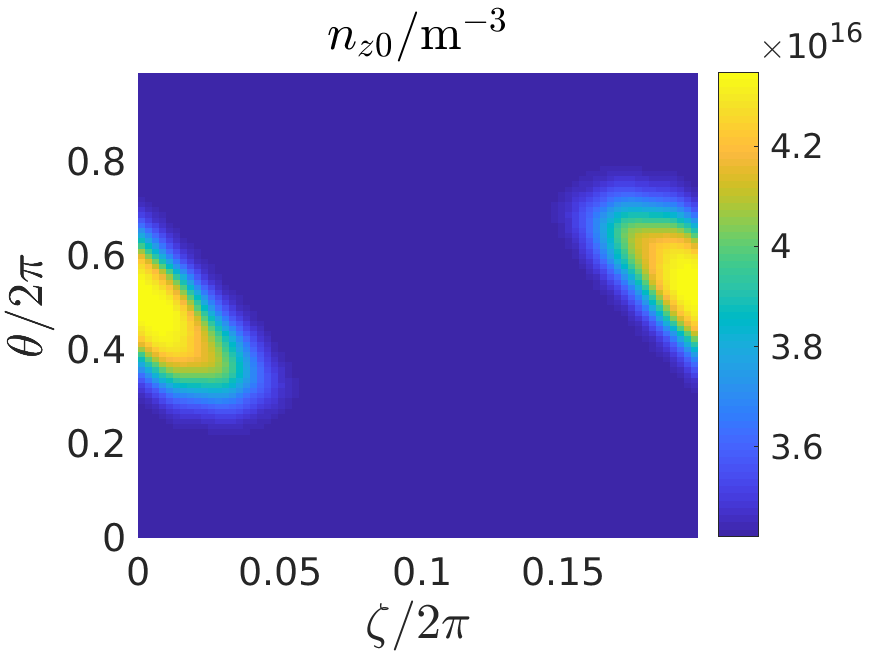}
    %\caption{$n_{z0}$}
  \end{subfigure}
  \caption{\small \label{fig:BPhinz0}  (left) A W7-X standard configuration vacuum field, with (middle) $\tilde{\Phi}$ and (right) $n_{z0}$, for $\tilde{\Phi}(B) = \A \e^{-(B-B_0)^2/(2\sigma^2)} - \C$ with $B_0 = B_\text{max}$, $\A=-10\,\mathrm{V}$ and $\sigma = 0.1|B_\text{max} - B_\text{min}|$, $Z=24$, $\lang n_z \rang = 3.472\times 10^{16}\,\rm{m^{-3}}$, and $T_i = T_z = 1\,\rm{keV}$. $\C$ is an integration constant set to make $\langle \tilde{\Phi} \rangle =0$.}
\end{figure}

To explore the implications of the flux-surface variation of $n_{z0}$ in \eqref{eq:DPhi}--\eqref{eq:DTi}, we consider a scenario where $\tilde{\Phi}$ is given by 
\begin{equation}
\tilde{\Phi}(B) = \A \e^{-(B-B_0)^2/(2\sigma^2)} - \C, \label{eq:PhiB}
\end{equation}
where $\A$ is the amplitude of the potential, $B_0$ is an extremum of $B$, $\sigma$ gives the width of $\tilde{\Phi}$, and $\C$ is an integration constant chosen to make $\langle \tilde{\Phi} \rangle =0$. The above $\tilde{\Phi}$ is intended to roughly emulate a potential perturbation due fully-circulating fast (collisionless) particles, although we are primarily interested in \eqref{eq:PhiB} as a simple test case, and will not be so concerned with whether it is a realistic fast-particle response.

We take $B$ from a Wendelstein 7-X vacuum field\footnote{We use a W7-X standard configuration at normalized radius $r_N = 0.6$, where $r_N = \sqrt{\psi_t/\psi_{t,\text{LCFS}}}$, with $\psi_t$ the toroidal flux and $\psi_{t,\text{LCFS}}$ its value at the last-closed flux-surface. The data is available at (Verified 2018-05-31) \\\url{https://github.com/landreman/sfincs/blob/master/equilibria/w7x-sc1.bc}}, and solve the magnetic differential equations for $u$ and $w$ numerically for this field. The magnetic field in Boozer coordinates (with $\zeta$, $\theta$ being toroidal and poloidal angle, respectively) is visualized in \autoref{fig:BPhinz0}, together with an example $\tilde{\Phi}$ and the resulting $n_{z0}$ for $Z=24$ and $\lang n_z \rang = 3.472\times 10^{16}\,\rm{m^{-3}}$.

To investigate the effects of a localized $n_z$ distribution, we performed a scan where the amplitude of the potential perturbation is increased. Specifically, the potential perturbation is centred at $B_{\text{max}}$ or $B_{\text{min}}$, and the amplitude $\A$ is scanned from $e\tilde{\Phi}/T_z = -0.1$ to $e\tilde{\Phi}/T_z = 0.1$ -- where a negative/positive sign corresponds to impurities accumulating/decumulating at $B_{\text{max}}$ or $B_{\text{min}}$. The ion temperature and density are $T_i = 1\,\rm{keV}$ and $n_i = 2\times 10^{20}\,\rm{m^{-3}}$; $m_i$ is taken as the proton mass. These values give $Z^2 \lang n_z \rang /n_i = 0.1$, so the impurities are trace.
For these parameters, the collisionalities are $\hat{\nu}_{ii} = 0.096$ and $\hat{\nu}_{zz} = 5.55$, where we have used $L_{\|} = (G + \iota I)/B_{00}$ as a proxy for the length-scale for parallel variations; here $\iota$ is the rotational transform, $G$ and $I$ are related to the magnetic field and defined in section 2.5 of \citet{helanderStellaratorReview2014}, $B_{00}$ is the $n=m=0$ Fourier-component of $B$ in Boozer coordinates.

\begin{figure}
  \centering
  \begin{subfigure}[b]{0.49\textwidth}
    \includegraphics[width=1\textwidth]{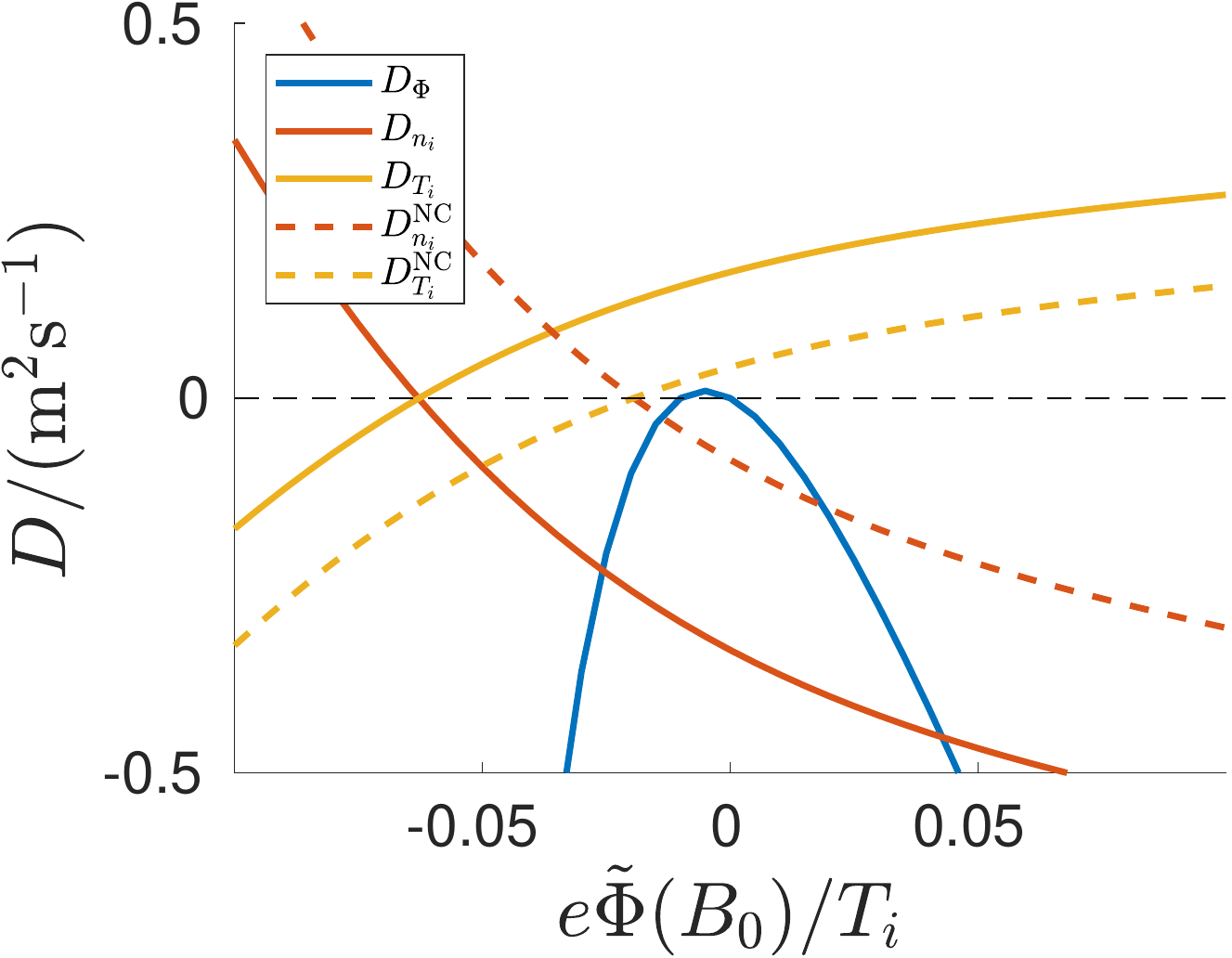}
    \end{subfigure}
    \begin{subfigure}[b]{0.49\textwidth}
      \includegraphics[width=1\textwidth]{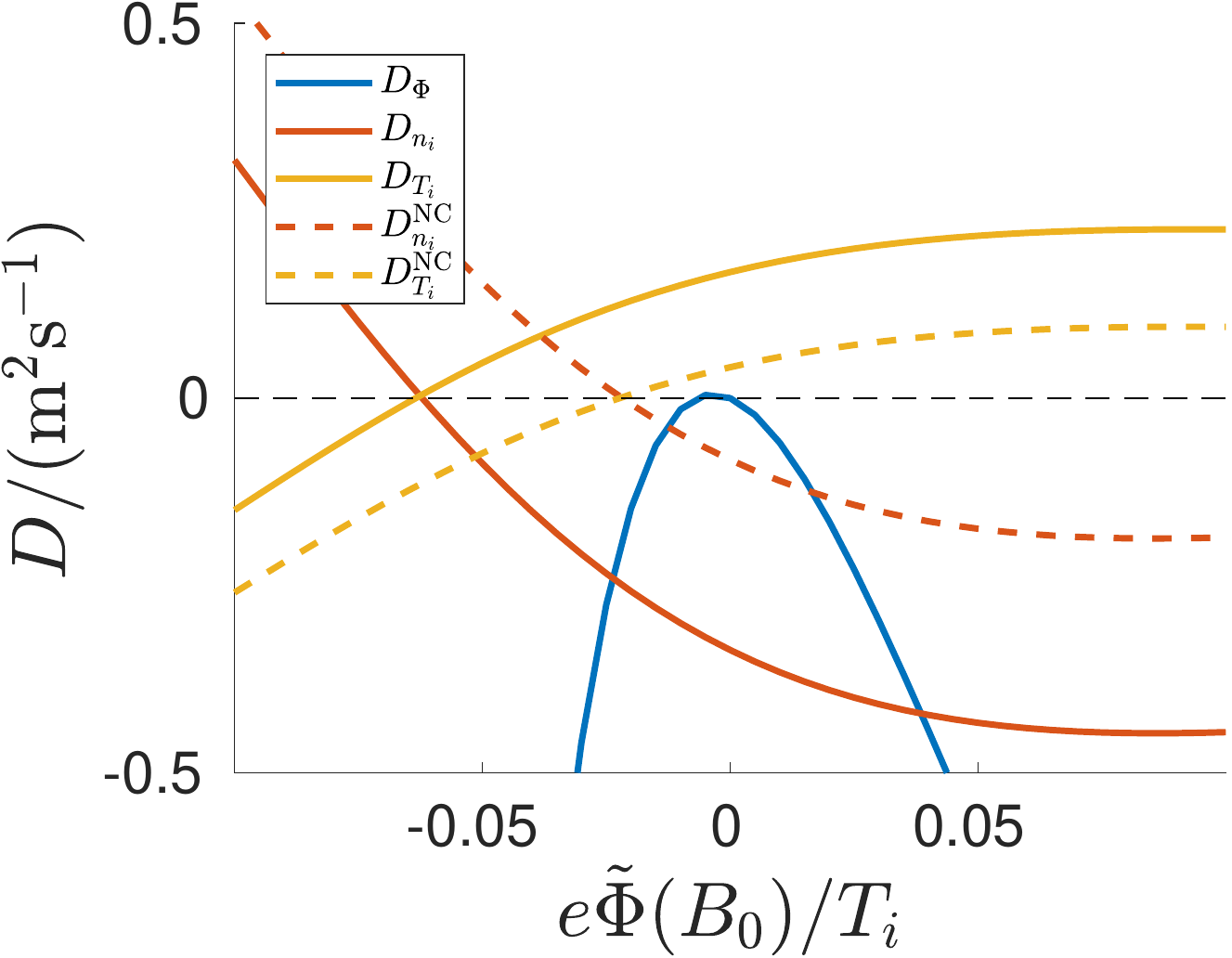}
    \end{subfigure}
    \caption{\label{fig:D_Phi1scan} Transport coefficients, for different potential perturbation amplitudes $\A$, for $\tilde{\Phi}$ localized around (left) $B_{\text{max}}$ and (right) $B_{\text{min}}$. $T_i = 1\,\rm{keV}$ and $n_i = 2\times 10^{20}\,\rm{m^{-3}}$, $\sigma = 0.1|B_\text{max} - B_\text{min}|$, $Z=24$, $\lang n_z \rang = 3.472\times 10^{16}\,\rm{m^{-3}}$; $D^{\text{NC}}$ refers to the neoclassical transport coefficients, while $D$ is the sum of classical and neoclassical.
At amplitudes roughly within $e\tilde{\Phi}/T_i \in  [-0.007,0]$, the flux due to an inward radial electric field is outward but very weak.}
\end{figure}

The resulting transport coefficients are shown in \autoref{fig:D_Phi1scan}. In the figures, $D$ (without superscript index) refers to the sum of neoclassical and classical $D$'s. For comparison, we also show $D^{\text{NC}}$; $D_{N_z}$ is not shown, since  $D_{N_z} = -D_\Phi - D_{n_i}$, and the Schwarz inequality causes it to always be non-negative, so that the question of whether impurities accumulate can be answered without its exact value. 
As indicated in \autoref{sec:class}, classical transport is dominant for this field configuration at $\tilde{\Phi}=0$, but we also see that the transport due to the radial electric field starts to dominate already at $e \A/T_i \sim 0.02$. 
When the radial electric field does not dominate, the impurities will be driven outwards when the temperature gradient is strong enough, i.e.\ we have \emph{temperature screening}.
Specifically, temperature screening occurs when $\d_\psi \ln T_i \geq -D_{n_i}\d_\psi \ln n_i/D_{T_i} \approx 2 \d_\psi \ln n_i$, and thus depends on the ratio $D_{n_i}/D_{T_i}$. This ratio is equal to $-2$ to within $1\%$ in the $\tilde{\Phi}$-amplitude window when radial electric field does not dominate, so the temperature screening condition is essentially unaffected, despite
the transport coefficients $D_{n_i}$ and $D_{T_i}$ varying by approximately $25\%$ in this window. 
We also see that both when $B_0 = B_{\text{max}}$ and $B_0 = B_{\text{min}}$, there is a very narrow amplitude range, roughly $e \A/T_i \in  [-0.007,0]$, in which the impurity flux due to an inward radial electric field is weakly positive.

%At higher potential amplitudes, the radial electric field starts to play a role, but it remains subdominant for realistic amplitudes $e\tilde{\Phi}/T_i \lesssim 0.05$.

From \autoref{fig:D_Phi1scan}, we also see that most of the variation in $D_{n_i}$ and $D_{T_i}$ comes from the neoclassical flux. This can partly be understood from the simpler form of the classical flux \eqref{eq:fluxc}, where the dependence on $n_z$ is linear, so that the localized $n_z$ perturbation due to $\tilde{\Phi}$ merely acts as a weight in the geometric factor $\lang n_z |\nabla \psi|^2/B^2\rang$, which here gives a small effect when integrated over the flux-surface. In contrast, the neoclassical flux \eqref{eq:fluxDnc} is non-linear in $n_z$, and the total flux through the flux-surface is set by a balance between inward and outward fluxes at different points on the flux-surface. 
% As the typical amplitudes in W7-X is  $e \tilde{\Phi}/T_i \sim 0.01$ \citep{garcia2017}, these effects are likely not of practical importance in this scenario.

%It should however be noted that the $D^{\text{NC}}$'s dependence on $n_{z0}$ is not linear, so an arbitrary $n_{z0}$ variation on the flux-surface cannot be viewed as a superposition of localized $n_{z0}$. These conclusions may thus not hold in general.

%Thus, strong localization is likely to lead to strong impurity accumulation if the radial electric field is pointing inwards, and temperature screening is insensitive to the amplitude of the $\Phi$ perturbation.

\begin{figure}
  \centering
  \begin{subfigure}[b]{0.49\textwidth}
    \includegraphics[width=1\textwidth]{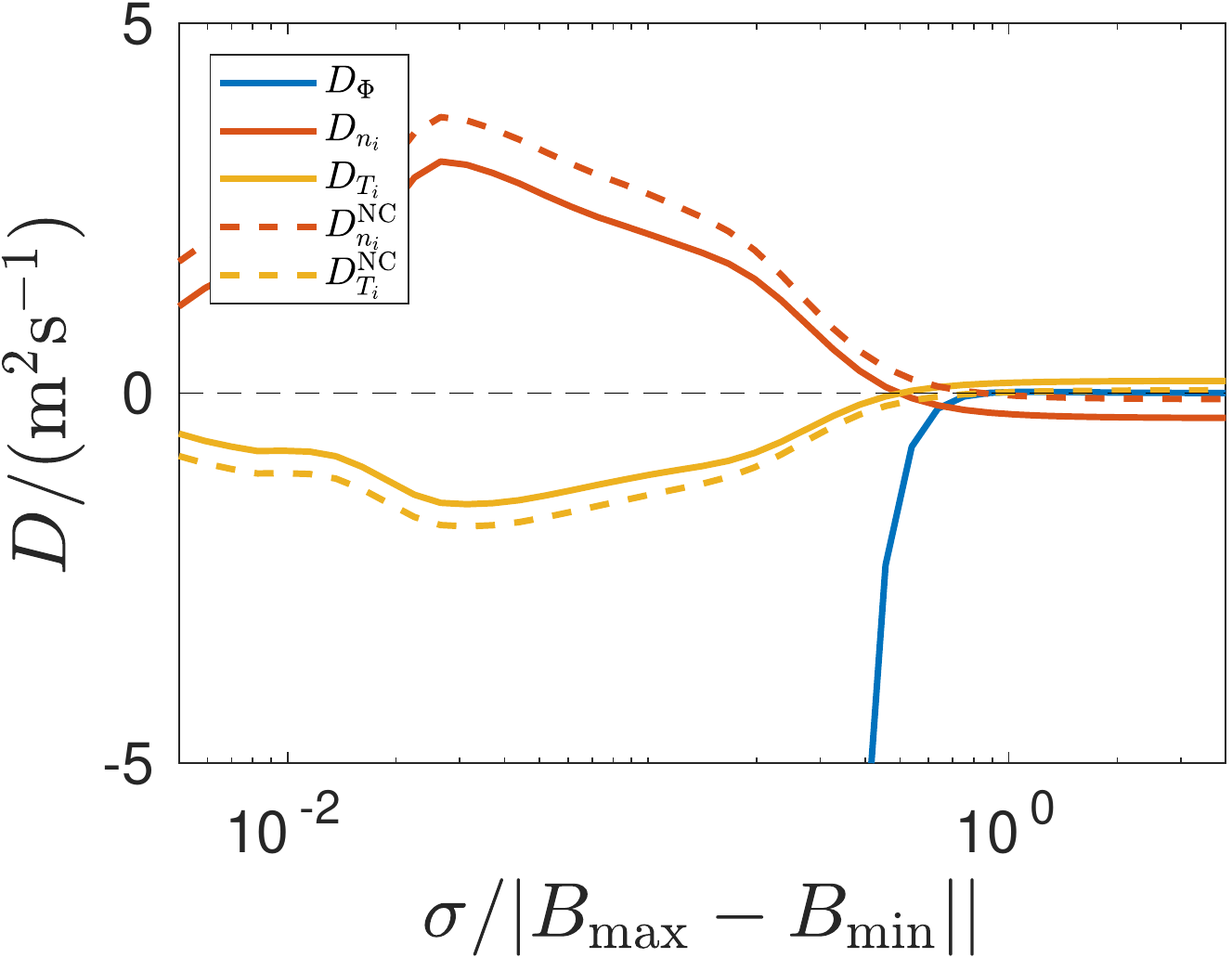}
  \end{subfigure}
\begin{subfigure}[b]{0.49\textwidth}
  \includegraphics[width=1\textwidth]{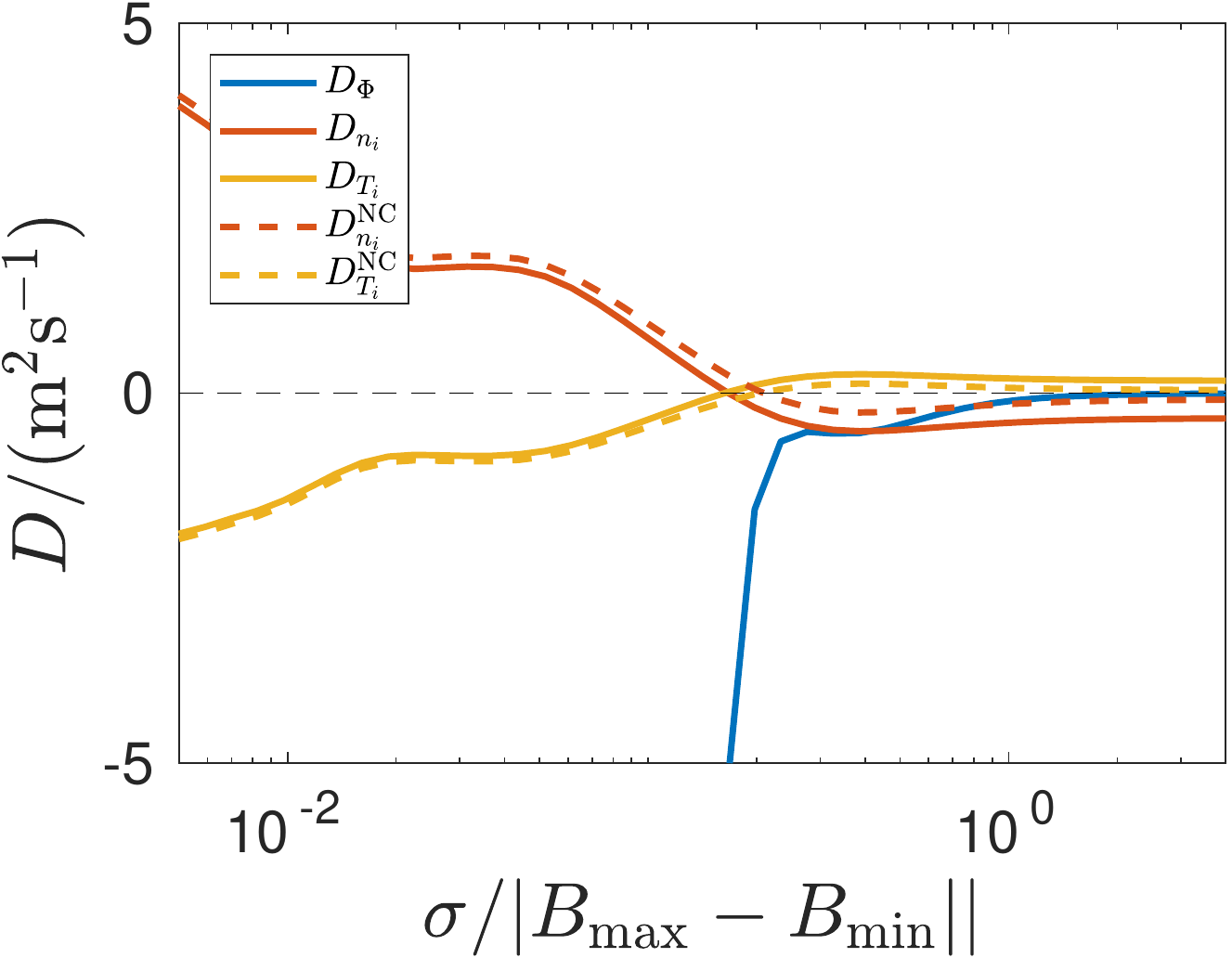}
  \end{subfigure}
  \caption{\label{fig:D_sigma_scan} Transport coefficient for varying $\sigma$, for $\tilde{\Phi}$ given by \eqref{eq:PhiB}, with $e\C/T_i= -0.025$ and $\A$ set to make $\langle \tilde{\Phi}\rangle =0$. The potential is centered around (left) $B_\text{max}$, and (right) $B_\text{min}$. Unless otherwise stated, quantities have the same value as in \autoref{fig:D_Phi1scan}.}
\end{figure}
%\todo{This part does not work nicely when the classical is included. The effect of only become significant when the amplitude is larger, but we then get numerical problems in $w$ for the smaller $\sigma$. Could exclude classical, but then we are studying an insignificant effect. (Using the LHD mock-up would likely work better.) Could exclude small $\sigma$, but then we do not get the homogeneous limit for small $\sigma$. }
%As the neoclassical flux is nonlinear, it may behave non-intuitively for different $n_z$.
To investigate the effects of more localized $n_{z0}$, we scanned the width of $\tilde{\Phi}$, while keeping $\C$ and $\langle n_{z0} \rangle$ constant. For small $\sigma$, this results in $n_{z0}$ that are very localized around the extremum of $B$.
The result is shown in \autoref{fig:D_sigma_scan}. From the figure, we see that $D_\Phi^{\text{NC}}$ diverges for localized $n_{z0}$. This is due to the $w_0^2$ terms in $D_\Phi^{\text{NC}}$:  $n_{z0} w_{0}$ obtained from \eqref{eq:w} is not localized to regions where $n_{z0}$ is localized, which results in a large $w_0$ where $n_{z0}$ is small. In contrast, the $D_{n_i}^{\text{NC}}$ and $D_{T_i}^{\text{NC}}$ remains finite, as $w_0$ only appears together with an $n_{z0}$ in those terms. In comparison to the neoclassical transport coefficients, the classical coefficients are only moderately affected by a more localized $n_{z0}$, for the same reasons as discussed in relation to the amplitude scan above.

% We see that for both wide and localized $\tilde{\Phi}$, the homogeneous limits are recovered. Note however that the smaller $\sigma$ gives a smaller integrated $\tilde{\Phi}$ as $\A$ is kept constant; when the integrated $\tilde{\Phi}$ is kept constant, high-amplitudes/small-widths give larger $D_\Phi$ -- again supporting the view that large $\tilde{\Phi}$ amplitudes result in large $D_\Phi$ and inward transport.

%Interestingly, the $D$'s cross their homogeneous values at around $\sigma=0.0104$ when $\tilde{\Phi}$ is centered around $B_\text{min}$.
%It should also be noted that the $D$'s dependence on $n_{z0}$ is not linear, so an arbitrary $n_{z0}$ variation on the flux-surface cannot be viewed as a superposition of localized $n_{z0}$.

\begin{figure}
  \centering
  \includegraphics[width=0.5\textwidth]{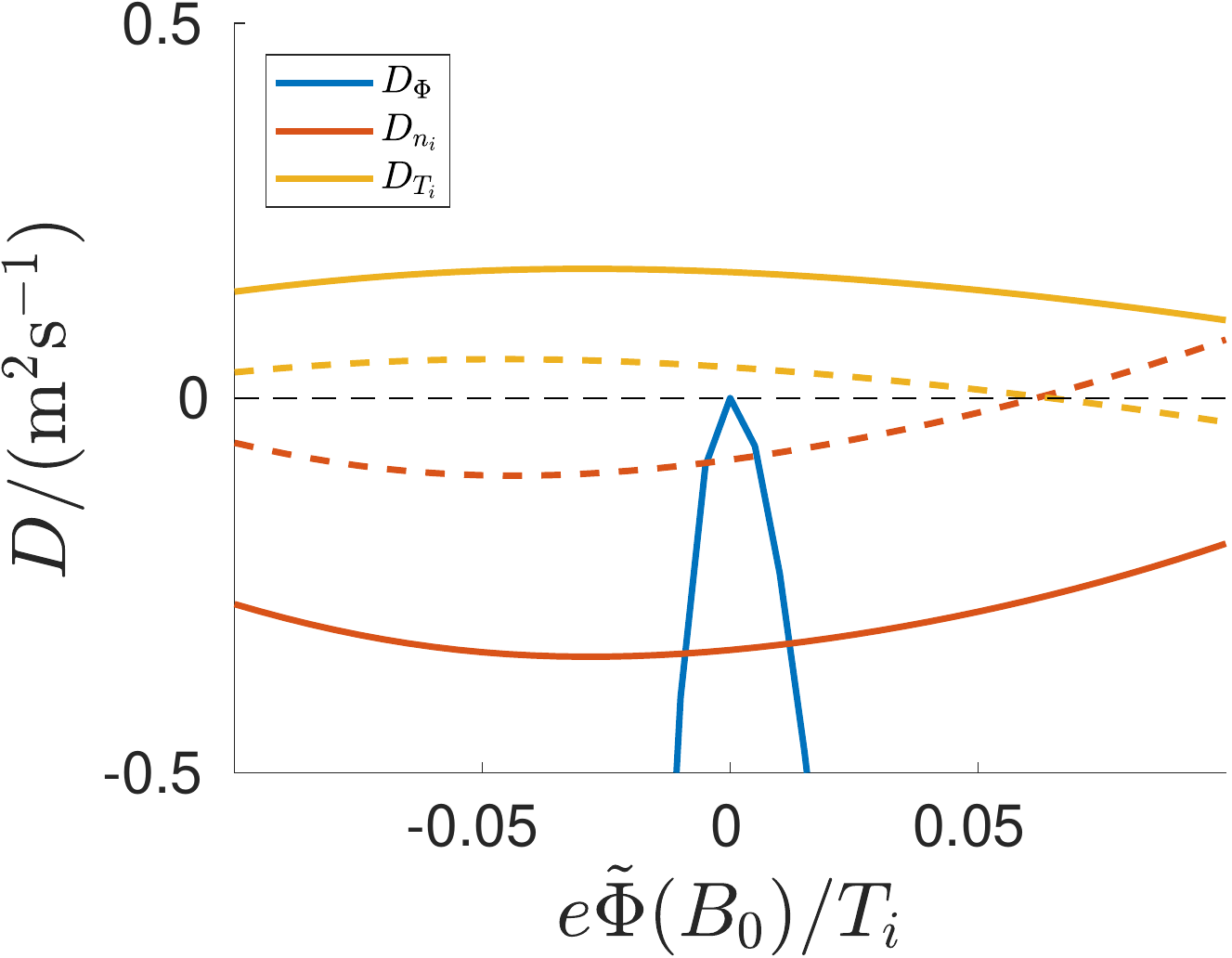}
  \caption{\label{fig:D_Phi1scan2} Figure corresponding to \autoref{fig:D_Phi1scan}, but with $n_{z0}$ concentrated around (or repelled from) the $B = (B_\text{min}/2 + B_\text{max}/2)$ contour.}
\end{figure}

%\todo{This part can be perhaps be replaced with more realistic distributions.}
To see whether this conclusion holds for more general $n_{z0}$, we let $B_0$ in \eqref{eq:PhiB} be a non-extremum point (within the flux surface), i.e.\  $B_0 \in (B_\text{min},B_{\text{max}})$. The resulting density distributions $n_{z0}$ will be concentrated or repelled from a contours of $B$, rather than points, and do not necessarily represent realistic density variations: rather, they provide simple test cases very different from those considered above, and thus give an indication of how general the above conclusions are.

For $B_0 = B_\text{min}/2 + B_\text{max}/2$, the resulting $D$'s are displayed in \autoref{fig:D_Phi1scan2}. Here, the $D_\Phi$ increases rapidly with the amplitude, and dominates the flux except for a small interval about $0$. Meanwhile, $D_{T_i}$ and $D_{n_i}$ are barely affected, with a slight reduction in magnitude when the impurities are repelled from $B_0$. 

\begin{figure}
  \centering
  \begin{subfigure}[b]{0.45\textwidth}
\includegraphics[width=1.0\textwidth]{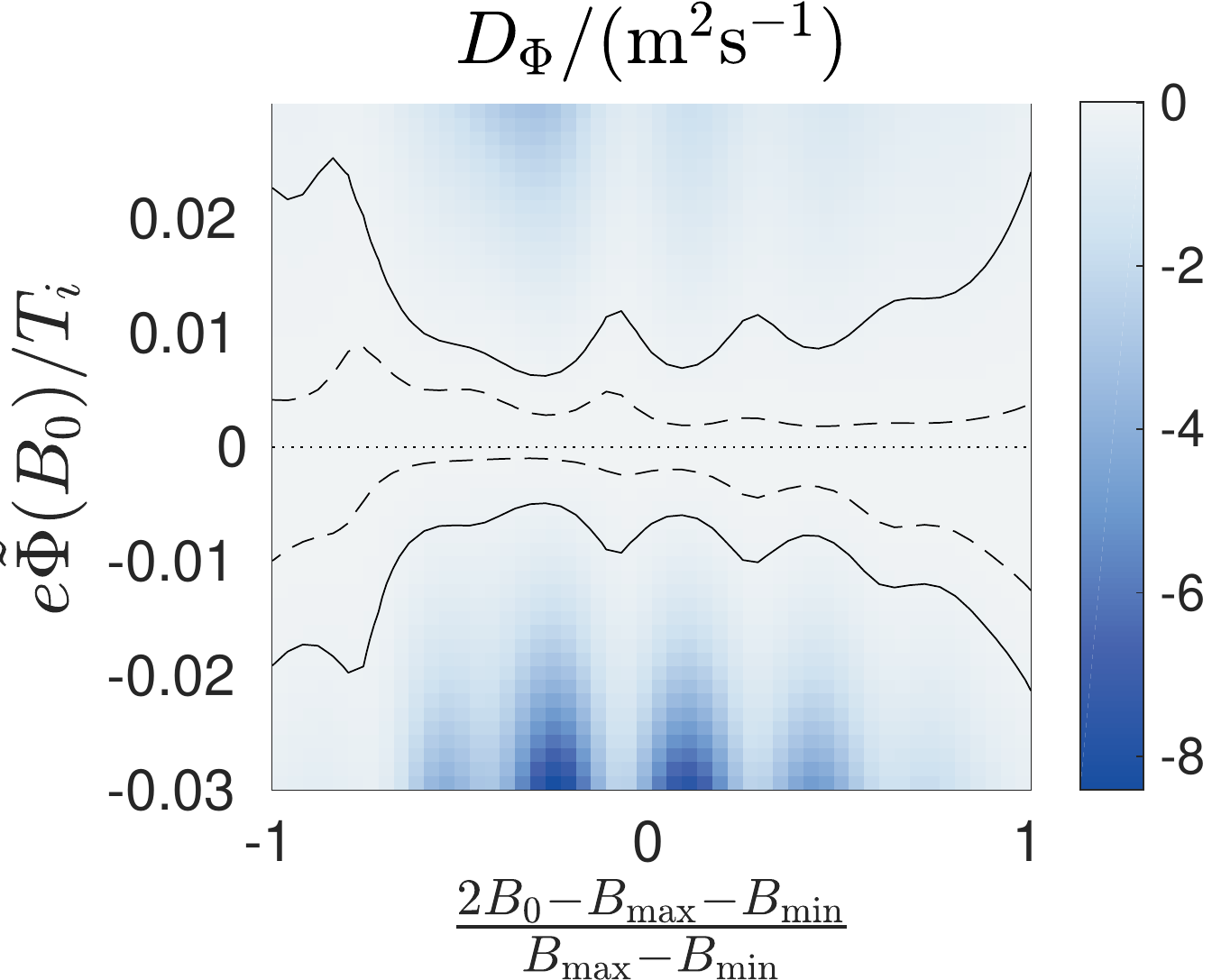}
    %\caption{$B$}
\end{subfigure}
\begin{subfigure}[b]{0.45\textwidth}
\includegraphics[width=1.0\textwidth]{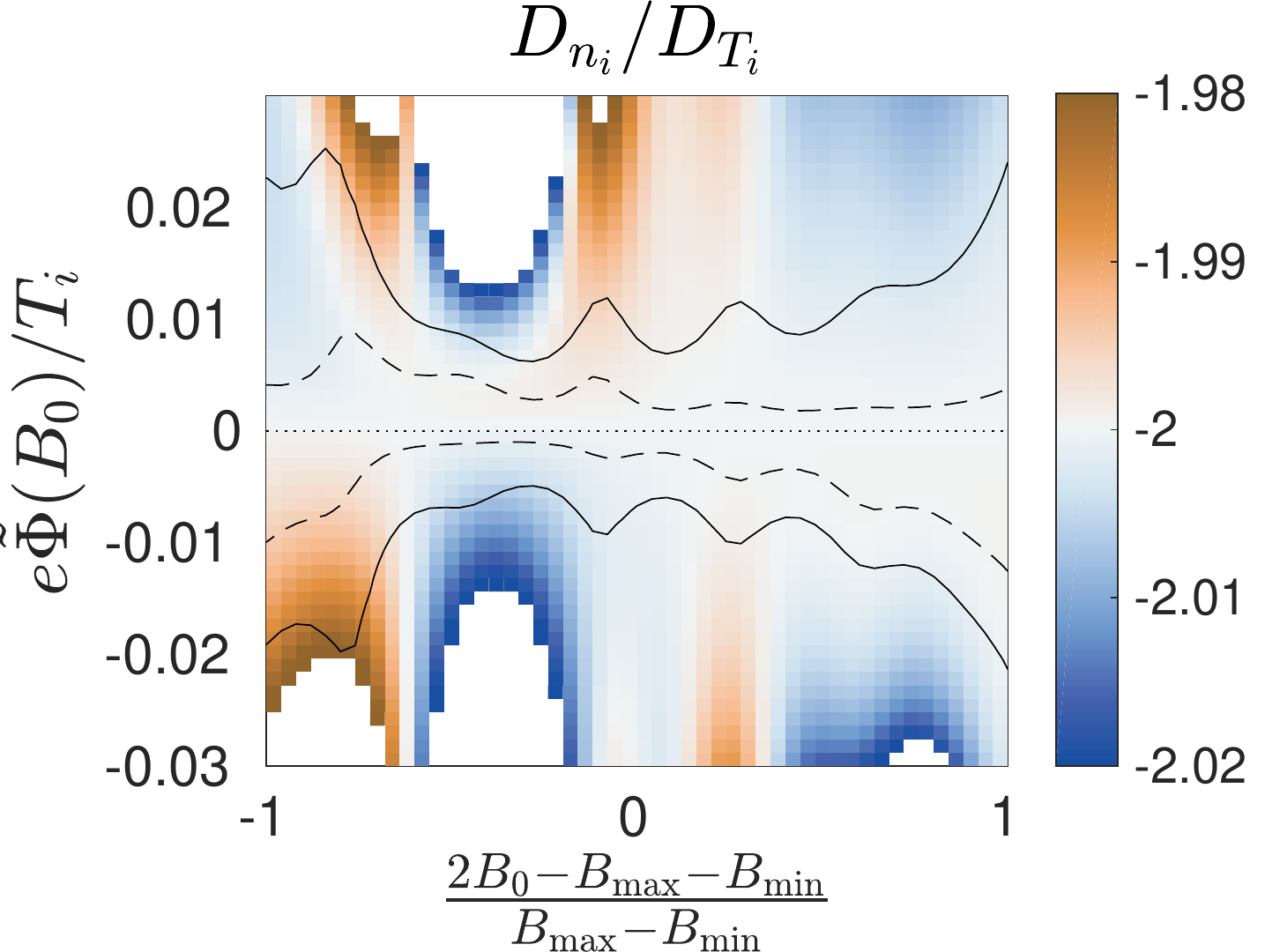}
    %\caption{$n_{z0}$}
  \end{subfigure}

\begin{subfigure}[b]{0.45\textwidth}
\includegraphics[width=1.0\textwidth]{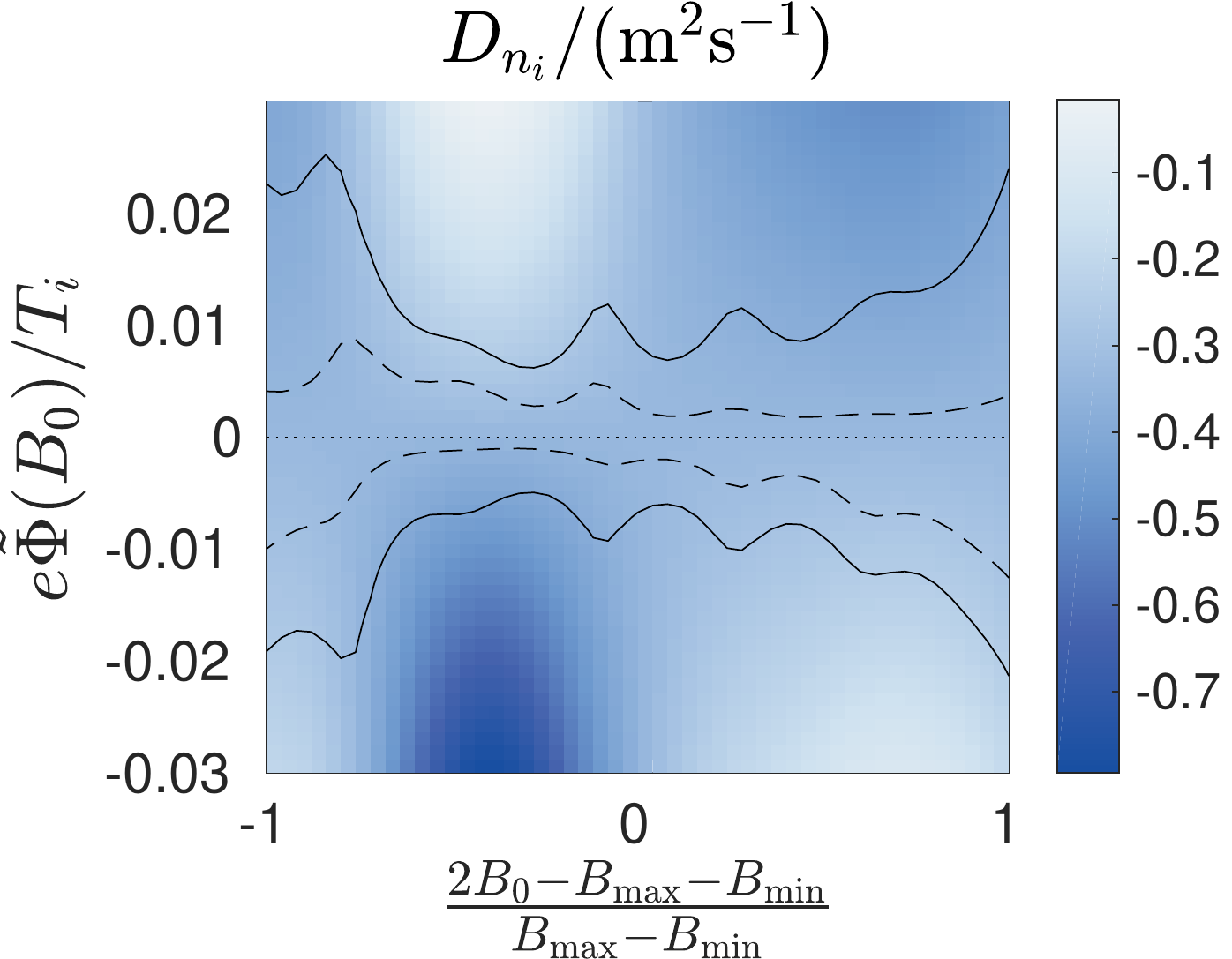}
%\caption{$\tilde{\Phi}$}
\end{subfigure}
\begin{subfigure}[b]{0.45\textwidth}
\includegraphics[width=1.0\textwidth]{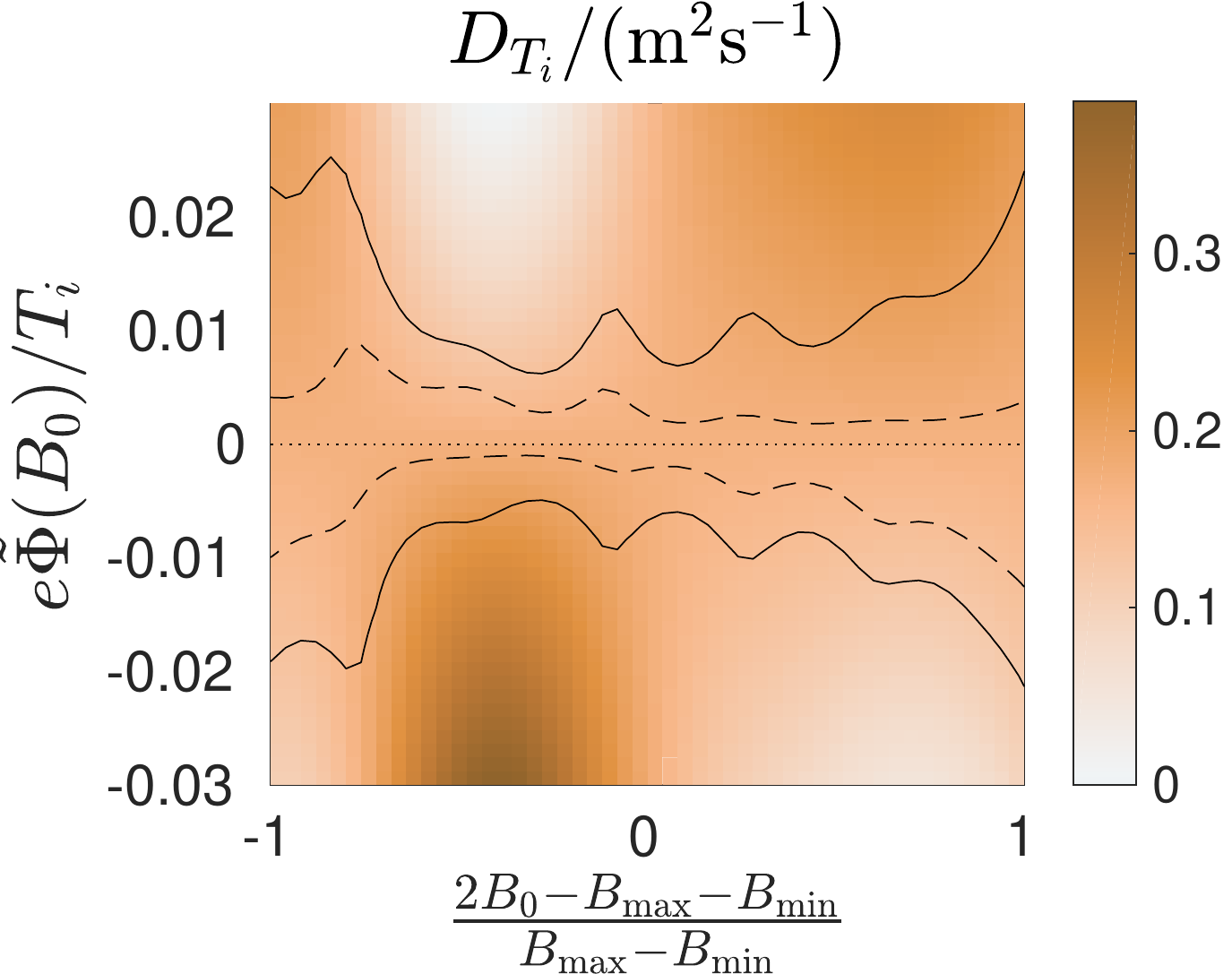}
    %\caption{$n_{z0}$}
\end{subfigure}
  \caption{\small \label{fig:D_B0scan}  Transport coefficient when $B_0$ is changed from $B_\text{min}$ to $B_\text{max}$. Apart from $B_0$, quantities have the same values as in \autoref{fig:D_Phi1scan}.
The black lines indicate potential amplitudes at which the radial electric field start to dominate: the solid line is where $D_\Phi =D_{n_i} + D_{T_i}$, the dashed is where $10D_\Phi =D_{n_i} + D_{T_i}$. {\bf Top-right figure:} $D_{n_i}/D_{T_i}$, which is within $1\%$ of $2$ for amplitudes where $D_{n_i} + D_{T_i} \geq D_\Phi$ (deviations larger than $1\%$ are white in the figure).} 
\end{figure}
To connect this result to when $B_0$ is an extremum, we scanned $B_0$ from $B_\text{min}$ to $B_\text{max}$. The resulting $D$'s are shown in \autoref{fig:D_B0scan}. 
Looking at $D_\Phi$, we see that as we go from $B_0 = B_\text{min}/2 + B_\text{max}/2$ ($x=0$ in the figure) towards the extrema ($x=1$ and $x=-1$), $D_\Phi$ tends to become less sensitive to the amplitude. 

From the top-right figure in \autoref{fig:D_B0scan}, we see that for all $B_0$, $D_{n_i}/D_{T_i}$ changes by less than $1\%$ within the amplitude interval where $D_{\Phi}$ is small enough to not notably affect temperature screening (this interval is within or slightly outside the dashed lines, which show where $10D_\Phi = D_{n_i} + D_{T_i}$).

To conclude, it thus appears that strong $\tilde{\Phi}$ perturbations are likely to lead to strong impurity accumulation if the radial electric field is pointing inwards, and that the condition for temperature screening is essentially unchanged from the $\tilde{\Phi} = 0$ case when the $\tilde{\Phi}$ perturbations are weak enough so that the radial electric field is not dominant.

% \begin{figure}
%   \centering
%   \begin{subfigure}[b]{0.32\textwidth}
% \includegraphics[width=1.0\textwidth]{{figs/u}.png}
%     %\caption{$B$}
%   \end{subfigure}
% \begin{subfigure}[b]{0.32\textwidth}
% \includegraphics[width=1.0\textwidth]{{figs/w0}.png}
% %\caption{$\tilde{\Phi}$}
%   \end{subfigure}
%   \caption{\small \label{fig:uw0}  $u$ and $w_0$ corresponding to the density and magnetic field in \autoref{fig:BPhinz0}.} 
% \end{figure}

\section{Summary \& conclusions}
We have derived expressions for the radial flux of high-$Z$ collisional impurities when the bulk ions are in the $1/\nu$ regime. In this limit, the impurity temperature is equilibrated with the bulk ions, while the impurity density can vary within the flux-surface. We have derived an expression for the parallel friction-force acting on the impurities, which can be used to solve for the impurity density variations on the flux-surface, given a mechanism for relating the impurity density to the electrostatic potential.

We considered in detail the trace impurity limit, with the impurity density set bu a Boltzmann response to an externally imposed the electrostatic potential.
Using simple models for $\tilde{\Phi}$ and a W7-X vacuum field, we have seen that large $\tilde{\Phi}$ amplitudes can cause the radial electric field to substantially contribute to the impurity transport, and lead to impurity accumulation when the radial electric field points inward. For smaller $\tilde{\Phi}$ amplitudes, temperature screening can be effective, and the condition for temperature screening is essentially the same as in the $\tilde{\Phi} = 0$ case, meaning that the temperature profile should be at least twice as steep as the density profile for screening to happen. In all cases, the contribution from classical transport is substantial, and even moderate $\tilde{\Phi}$ can cause the electric field to dominate if classical transport is not accounted for.

It is however not straightforward to extrapolate from these results to general $\tilde{\Phi}$, as the neoclassical impurity flux does not depend linearly on $n_z$, so the flux from a general $n_z$ flux-surface distribution is not a superposition of fluxes from simpler $n_z$ distributions. Realistic $\tilde{\Phi}$ or $n_z$ distributions may be needed to evaluate the fluxes accurately, especially when neoclassical transport is comparable to or stronger than the classical. We refer the interested reader to \citet{calvo2018}, where an LHD equilibrium with a $\tilde{\Phi}$ set up by particle trapping effects of the bulk ions is considered. 

\section*{Acknowledgement} The authors are grateful for the fruitful discussions with the authors of \citet{calvo2018}, who considered the same problem independently and in parallel to us, from which we have benefited. %After taking part in their results, we introduced $w$ and thus obtained a compact flux-friction relation and diffusion coefficients.
% \todo{Thanks to whoever calculated the W7-X vacuum field? Anyone else?}
This work has been carried out within the framework of the EUROfusion
Consortium and has received funding from the Euratom research and training
programme 2014-2018 under grant agreement No 633053. The views and
opinions expressed herein do not necessarily reflect those of the European
Commission.
SB and IP were supported by the International Career Grant of Vetenskapsr{\aa}det (Dnr.~330-2014-6313) and IP by Marie Sklodowska Curie Actions, Cofund, Project INCA 600398. SB's visit to Greifswald was supported by Chalmersska forskningsfonden. %The simulations used SNIC computational resources at Kebnekaise (Dnr: 2017/3-29) and Hebbe (Dnr: 2017/1-95).

\appendix

\section{Solving the ion drift-kinetic equation}
\label{sec:sidke}

In this section, we solve the ion kinetic equation \eqref{eq:fi1m1}--\eqref{eq:fi11}  for $F_{i1(-1)}$ and $F_{i1(0)}$. The solution follows \nouncite{newton2017}, but here $n_z$ is allowed to vary on the flux-surface.
Since we assume $e\tilde{\Phi}/T_i \sim Z^{-1}$, the potential energy of the bulk ions is approximately constant on the flux-surface, and we change variables from $\E$, $\mu$ to the approximate invariants $v$ and $\lambda$.

We note that since we only use $F_{i1}$ to calculate the ion-impurity friction force, we only need the part of $F_{i1}$ that is odd in $v_{\|}$. We thus split \eqref{eq:fi1m1}--\eqref{eq:fi11} into odd and even equations.

Denoting the odd (even) part of the distribution function with a minus (plus) superscript, the order $\hat{\nu}^{-1}$ equations become
\begin{align}
  &v_\| \nabla_\| F_{i1(-1)}^+  = 0\label{eq:ofi1m1} \\
  &v_\| \nabla_\| F_{i1(-1)}^-  = 0\label{eq:efi1m1},
\end{align}
which simply states that $F_{i1(-1)}$ is constant along field-lines,
\begin{equation}
F_{i1(-1)} = F_{i1(-1)}(l_0),
\end{equation}
where $l_0$ is an arbitrary point on the field line. In the trapped region, this implies that $F_{i1(-1)}^- = 0$, since it must vanish at bounce-points. In the passing region, $F_{i1(-1)}(l_0)$ is set by solvability conditions to the next order equations.

To order $\hat{\nu}^0$, we have that
\begin{align}
  &v_\| \nabla_\| F_{i1(0)}^+ = C_{i}^-[F_{i1(-1)}]. \label{eq:ofi10} \\
  &v_\| \nabla_\| F_{i1(0)}^- + \vec{v}_d \cdot \nabla f_{i0} = C_{i}^+[F_{i1(-1)}]. \label{eq:efi10} 
\end{align}
In the passing region, the odd and even part of $F_{i1(-1)}(l_0)$ can be determined by acting with $\lang \frac{B}{v_\|} \dots \rang$ on  equations \eqref{eq:ofi10}--\eqref{eq:efi10}, resulting in
\begin{align}
  &\lang \frac{B}{v_\|} C_{i}^-[F_{i1(-1)}] \rang = 0 \label{eq:condm1}\\
  &\lang \frac{B}{v_\|}  C_{i}^+[F_{i1(-1)}] \rang = \lang \frac{B}{v_\|} \vec{v}_d \cdot \nabla \psi  \rang \p_\psi f_{i0} = 0, \label{eq:evencondm1}
\end{align}
where the latter equality follows from writing $\vec{v}_d \cdot \nabla \psi= v_\| (\vec{b} \times \nabla \psi) \cdot \nabla \left(\frac{v_\|}{\Omega_i}\right)$.
The odd and even parts of the collision operator are
\begin{align}
  C_i^+[X] &= (\nu_{ii}^D + \nu_{iz}^D) \mathcal{L}X^+ \\
  C_i^-[X] &= (\nu_{ii}^D + \nu_{iz}^D) \mathcal{L}X^- + \frac{m_i f_{i0}}{T_i} v_\|(\nu_{ii}^D  U_\| + \nu_{iz}^DV_{z\|}), \label{eq:Cminus}
\end{align}
with $\mathcal{L} = \frac{2v_\|}{v^2B} \frac{\p}{\p \lambda} \lambda v_\| \frac{\p}{\p \lambda}$. \autoref{eq:condm1} implies that $F_{i1(-1)}^+$ is constant in $\lambda$, so that $C_{i}^+[F_{i1(-1)}]=0$ in the passing region. The same argument applies to $F_{i1(-1)}^-$, unless there is a parallel impurity flow in \eqref{eq:Cminus} to order $\hat{\nu}^{-1}$ to act as a source in \eqref{eq:condm1}. Such order $\hat{\nu}^{-1}$ flows cannot arise in the mixed-collisionality regime, so $F_{i1(-1)}^-=0$ \citep{newton2017}. However, to make the $F_{i1}$ formulas in this section apply for any impurity collisionality, we will nevertheless allow for $F_{i1(-1)}^-\neq 0$ below, as it turns out to not be inconvenient to calculate $F_{i1(-1)}^-$ together with $F_{i1(0)}^{-}$.
%We will for now assume that there is no order $\hat{\nu}^{-1}$ flow, so that $F_{i1(-1)}^- = 0$, but it will become apparent that this assumption is 

To solve for $F_{i1(0)}^{-}$, we note that \eqref{eq:efi10} can be formally solved by integrating along a field-line; using $l$ to denote the distance along the field line, we have
\begin{equation}
F_{i1(0)}^{-}(l) =F_{i1(0)}^{-}(l_0) +\int\limits_{l_0}^{l} \frac{dl'}{v_\|} \left[C^+[F_{1(-1)}]-\vec{v}_d \cdot \nabla f_{i0}(l')\right], \label{eq:fi1}
\end{equation}
where the integration constant $F_{i1(0)}^{-}(l_0)$ again is set by the solvability condition of the next-order equation. 
Taking $l_0$ to be a bounce-point, $B(l_0) = 1/\lambda$, we have that $F_{i1(0)}^{-}(l_0)=0$ in the trapped region. 
To determine $F_{i1(0)}^{-}(l_0)$ the passing region, we again act with $\lang \frac{B}{v_\|} \dots \rang$ on the next-order odd equation, which gives
\begin{equation}
\lang \frac{B}{v_\|} C_i^-[F_{i1(0)}]\rang = 0. \label{eq:cond}
\end{equation}
Note that this is essentially the same equation as \eqref{eq:condm1}. Thus, the total distribution $F_{i1} \approx F_{i1(0)} + F_{i1(-1)}$ can be written on the same form as $F_{i1(0)}$, but with an order $\hat{\nu}^{-1}$ contribution to the integration constant $F_{i1}^{-}(l_0) \approx F_{i1(-1)}^{-}(l_0) + F_{i1(0)}^{-}(l_0)$. As such, it is in some sense irrelevant whether parts of $V_{z\|}$ are order $\hat{\nu}^{-1}$ or $\hat{\nu}^{0}$, as $F_{i1(0)} + F_{i1(-1)}$ is not affected by the way this decomposition of $V_{z\|}$ is done.
%but the impurity flmakes the distribution non-trivial.
% Defining the total distribution function and integration constant
% \begin{equation}
% F_{i1}^- = F_{i1(0)}^-, \qquad F_{i1}^{-}(l_0) = F_{i1(-1)}^{-}(l_0) + F_{i1(0)}^{-}(l_0), \label{eq:totalF1}
% \end{equation}
% \eqref{eq:condm1} and \eqref{eq:cond} can be summed to yield an equation for $F_{i1}^{-}(l_0)$
% To have $l_0$ be continuous when going from trapped- to passing-region, we choose $l_0$ in the passing region so that $B(l_0) = B_{\text{max}}$.

Inserting \eqref {eq:Cminus} into \eqref{eq:cond} gives the following equation for the integration constant $F_{i1}^{-}(l_0) = F_{i1(-1)}^{-}(l_0) + F_{i1(0)}^{-}(l_0)$
\begin{align}
  &\frac{\p}{\p \lambda} F_{i1}^{-}(l_0)  \label{eq:intC}
\\
=&  - \frac{mv^2}{2 \lang\left[1 + \frac{\nu_{iz}^D(l)}{\nu_{ii}^D}\right] v_\| \rang } \left( \frac{1}{e} \lang \left[1 + \frac{\nu_{iz}^D(l)}{\nu_{ii}^D}\right]  g_4(l,\lambda) \rang \frac{\p f_{i0}}{\p \psi}
+  \frac{1}{T_i} f_{i0} \lang B \left[U_\| + \frac{\nu_{iz}^D(l)}{\nu_{ii}^D}V_{z\|}\right] \rang \right), \nonumber
\end{align}
in the passing region. 
To account for the $\vec{v}_d \cdot \nabla f_{i0}$ term, we have introduced the geometric function \citep{nakajima1989}
\begin{equation}
g_4(\lambda,l) = v_\| \int_{l_0}^l \d l' \left(\vec{b} \times \nabla \psi\right) \cdot \nabla \left(\frac{1}{v_\|}\right).
\end{equation}
Note that $C_{i}^+[F_{i1(-1)}] =0$ in this region. In the trapped region, on the other hand, $F_{i1}^{-}(l_0)=0$ but $C_{i}^+[F_{i1(-1)}] \neq 0$. However, the $C_{i}^+[F_{i1(-1)}]$-term nevertheless gives no contribution to the parallel flow or friction force in this region \citep{helanderParraNewton2017}.

\section{Parallel friction force}
\label{sec:b}
Once $U_\|$ and $V_{z\|}$ are known, we can use \eqref{eq:fi1} and \eqref{eq:intC} to directly evaluate the parallel friction force acting on the impurities. From our mass-ratio expanded ion-impurity collision operator \eqref{eq:ciz} and the self-adjointness of the Lorentz operator, we have
\begin{equation}
  \begin{aligned}
    R_{iz\|} =  &\int \d^3 v m_i v_\| \nu_{iz}^D(v) \left( \frac{m_i v_\| V_{z\|}}{T_i} f_{i0} - F_{i1}^- \right) \\
    = & \frac{n_i m_i}{\tau_{iz}} \left(V_{z\|} - \frac{T_i}{e} \left[ A_{i1} - \frac{3}{2} A_{i2}\right] Bu - BP(\psi) \right), 
  \end{aligned}
\end{equation}
where $u$ satisfies the magnetic equation \eqref{eq:u} and $P$ is a flux-function which contains the contribution from the integration constant $F_{i1}^{-}(l_0)$
\begin{equation}
P(\psi) \equiv \frac{\tau_{iz}}{Bn_i} \int \d^3 v  v_\| \nu_{iz}^D(v) F_{i1}^{-}(l_0). \label{eq:P1}
\end{equation}
$P(\psi)$ can be evaluated using \eqref{eq:intC} and partial integration in $\lambda$
\begin{equation}
P(\psi) =
 \frac{\lang B U_{\|}\rang }{\lang B^2 \rang} b_1 +  \frac{\lang \alpha B V_{z\|} \rang }{\lang B^2 \rang} b_4 + \frac{T_i}{e}\frac{A_{1i} - \frac{5}{2} A_{2i}}{\lang B^2 \rang} b_2 + \frac{T_i}{e}\frac{A_{2i}}{\lang B^2 \rang} b_3 \label{eq:P2}
\end{equation}

% \begin{equation}
% \begin{aligned}
% P(\psi) = & \frac{m_i \pi }{n_i} 
%    \int_0^\infty v^5 \tau_{iz}\nu_D^{iz}\int\limits_0^{B_\text{max}^{-1}} \d\lambda \lambda 
% \frac{1}{\lang\left[1 + \frac{\nu_{iz}^D(l)}{\nu_{ii}^D}\right] v_\| \rang } \left( \frac{1}{e} \lang \left[1 + \frac{\nu_{iz}^D(l)}{\nu_{ii}^D}\right]  g_4(l,\lambda) \rang \frac{\p f_{i0}}{\p \psi} \right.
% \\
% &\left.+  \frac{1}{T_i} f_{i0} \left[\lang B U_\| \rang + \lang \frac{\nu_{iz}^D(l)}{\nu_{ii}^D}BV_{z\|} \rang \right] \right).
% \end{aligned}
% \end{equation}
where we have introduced
\begin{align}
b_1 &= \frac{m \pi \lang B^2 \rang }{n_i T_i \{\nu_{D}^{iz} \}}
\int_0^\infty \!\d v  v^4
\nu_D^{iz} f_{i0}  \int_0^{1/B_\text{max}} \!\d \lambda\,  \lambda 
\frac{1}{\lang \xi \rang +  \lang \frac{\nu_{iz}^D}{\nu_{ii}^D}  \xi \rang } \label{eq:b1}\\
b_2 &= \frac{m \pi \lang B^2 \rang }{n_i T_i \{\nu_{D}^{iz} \}}
\int_0^\infty \!\d v  v^4
\nu_D^{iz} f_{i0} \int_0^{1/B_\text{max}} \!\d \lambda\,  \lambda 
\frac{\left[ \lang g_4 \rang + \lang \frac{\nu_{iz}^D}{\nu_{ii}^D}  g_4  \rang \right]}{\lang \xi \rang + \lang  \frac{\nu_{iz}^D}{\nu_{ii}^D} \xi \rang }   \label{eq:b2}\\  
b_3 &= \frac{m \pi \lang B^2 \rang }{n_i T_i \{\nu_{D}^{iz} \}}
\int_0^\infty \!\d v  v^4 \frac{m_iv^2}{2T_i}
\nu_D^{iz} f_{i0} \int_0^{1/B_\text{max}} \!\d \lambda\,  \lambda 
\frac{\left[ \lang g_4 \rang + \lang \frac{\nu_{iz}^D}{\nu_{ii}^D}  g_4  \rang \right]}{\lang \xi \rang + \lang \frac{\nu_{iz}^D}{\nu_{ii}^D} \xi \rang } \label{eq:b3}\\
b_4 &= \frac{m \pi \lang B^2 \rang }{Z^2 n_z T_i \{\nu_{D}^{iz} \}}
\int_0^\infty \!\d v  v^4
\nu_D^{iz} \frac{\nu_D^{iz}}{\nu_{ii}^D} f_{i0} \int_0^{1/B_\text{max}} \!\d \lambda\,  \lambda 
\frac{1}{\lang \xi \rang + \lang \frac{\nu_{iz}^D}{\nu_{ii}^D}  \xi \rang },\label{eq:b4}
\end{align}
with $\xi = \vec{v} \cdot \vec{b}/v$. The velocity average $\{ \cdot \}$ is defined as
\begin{equation}
\{F(v) \} \equiv \frac{8}{3\sqrt{\pi}} \int_0^\infty \d x\,  F(x) x^4  e^{-x^2},
\end{equation}
where $x = v/v_{Ti}$.

To have the boundary terms from the partial integration disappear in \eqref{eq:P2}, we have defined $l_0$ through $B(l_0) = B_\text{max}$. This makes our choice of $l_0$ continuous when going from the trapped to the passing region, and thus ensures that $F_{i1}^{-}(l_0)$ is zero at the trapped-passing boundary $\lambda = 1/B_\text{max}$.

\section{Momentum restoring term $U_\|$}
\label{sec:U}
The momentum restoring term in the ion-ion model collision operator \eqref{eq:cii} is calculated so that ion-ion collisions conserve momentum. Specifically, we have
\begin{equation}
U_\| = \frac{1}{n_i \{\nu_{D}^{ii} \}} \int \d^3 v v_\| \nu_D^{ii}  F_{i1}^{-}.
\end{equation}
Inserting $F_{i1}^{-}$ from \eqref{eq:fi1} and using \eqref{eq:intC}, we get
\begin{equation}
  \begin{aligned}
   & \lang B U_{\|} \rang (1 - a_1) \\=& \frac{T_i }{e} \left(\left[a_2 + \lang u B^2 \rang\right] A_{1i} + \left[a_3 - \frac{5}{2} a_2 - \lang u B^2 \rang \eta\right] A_{2i}\right) + a_4 \lang \alpha B V_{z\|} \rang.
  \end{aligned}
  \label{eq:BU2}
\end{equation}
where we have defined the geometry-impurity dependent flux-surface constants
\begin{align}
a_1 &= \frac{m \pi \lang B^2 \rang }{n_i T_i \{\nu_{D}^{ii} \}}
\int_0^\infty \!\d v  v^4
\nu_D^{ii} f_{i0}  \int_0^{1/B_\text{max}} \!\d \lambda\,  \lambda 
\frac{1}{\lang \xi \rang +  \lang \frac{\nu_{iz}^D}{\nu_{ii}^D}  \xi \rang } \label{eq:a1}\displaybreak[0]\\
a_2 &= \frac{m \pi \lang B^2 \rang }{n_i T_i \{\nu_{D}^{ii} \}}
\int_0^\infty \!\d v  v^4
\nu_D^{ii} f_{i0} \int_0^{1/B_\text{max}} \!\d \lambda\,  \lambda 
\frac{\left[ \lang g_4 \rang +  \lang \frac{\nu_{iz}^D}{\nu_{ii}^D}  g_4  \rang \right]}{\lang \xi \rang + \lang  \frac{\nu_{iz}^D}{\nu_{ii}^D}  \xi \rang }   \label{eq:a2}\displaybreak[0]\\  
a_3 &= \frac{m \pi \lang B^2 \rang }{n_i T_i \{\nu_{D}^{ii} \}}
\int_0^\infty \!\d v  v^4 \frac{m_iv^2}{2T_i}
\nu_D^{ii} f_{i0} \int_0^{1/B_\text{max}} \!\d \lambda\,  \lambda 
\frac{\left[ \lang g_4 \rang +\lang  \frac{\nu_{iz}^D}{\nu_{ii}^D}  g_4  \rang \right]}{\lang \xi \rang +  \lang \frac{\nu_{iz}^D}{\nu_{ii}^D} \xi \rang } \label{eq:a3}\displaybreak[0]\\
a_4 &= \frac{m \pi \lang B^2 \rang }{Z^2 n_z T_i \{\nu_{D}^{ii} \}}
\int_0^\infty \!\d v  v^4
\nu_{iz}^D f_{i0} \int_0^{1/B_\text{max}} \!\d \lambda\,  \lambda 
\frac{1}{\lang \xi \rang + \lang  \frac{\nu_{iz}^D}{\nu_{ii}^D} \xi\rang }. \label{eq:a4}
\end{align}

\section{Solvability condition and $K_z$}
\label{sec:solvcond}
\autoref{eq:Gpar1} specifies $V_{z\|}$ up to a flux-function $K_z$ (c.f.\ \eqref{eq:Gpar2}). This $K_z$ can be determined from solvability condition of \eqref{eq:parallel}, which states that
\begin{equation}
\lang \frac{BR_{z\|}}{n_z}\rang = 0.
\end{equation}
Inserting \eqref{eq:pFF}, the solvability condition becomes
\begin{equation}
  \begin{aligned}
    \frac{n_i m_i}{n_z \tau_{iz}} \left(\lang B V_{z\|} \rang - \frac{T_i}{e} \left[ A_{i1} - \frac{3}{2} A_{i2}\right] \lang uB^2 \rang - \lang B^2 \rang P(\psi) \right) = 0. \label{eq:pFF2}
  \end{aligned}
\end{equation}

In the $\Delta \ll 1$ limit, we can insert our expression for $V_{z\|}$, \eqref{eq:Gpar2}, to solve for $K_z$. This results in \eqref{eq:Kz1}, where we have defined
\begin{align}
  c_1 &= b_1 + a_1 c_1 \implies c_1 = b_1/(1-a_1)\label{eq:c1},\\
  c_2 &= b_2 + a_2 c_1 \label{eq:c2}, \\
  c_3 &= b_3 + a_3 c_1 \label{eq:c3}, \\
  c_4 &= b_4 + a_4 c_1 \label{eq:c4},
\end{align}
for the sake of compactness.

\section{Trace impurity limit of some expressions}
\label{sec:abtrace}
In the trace impurity limit, $\alpha \equiv Z^2 n_z/n_i \ll 1$, the $a_i$, $b_j$ and $c_k$'s simplify considerably, yielding an expression for $K_z$ in terms of standard geometry functions. Specifically,
\begin{align}
a_1 & = b_1 = f_c, \label{eq:tra1}\displaybreak[0]\\
a_2 &= b_2 = f_s,  \displaybreak[0]\\  
  a_3 &=  f_s \left(\frac{5}{2} - \eta\right), \displaybreak[0]\\
  b_3 &=  f_s,  \displaybreak[0]\\
  a_4 & = \frac{f_c}{\alpha}\frac{\{\nu_{D}^{iz}\}}{\{\nu_{D}^{ii} \}}, \displaybreak[0]\\
  b_4 &= \frac{f_c}{\alpha}\frac{\{{\nu_D^{iz}}^2/\nu_D^{ii} \}}{\{\nu_{D}^{iz} \}}. \label{eq:trb4}
\end{align}
Note that $a_4$ and $b_4$ only appear in terms containing $\alpha$, which are negligible in the trace-limit. Here,
\begin{align}
f_c &= \frac{3 \lang B^2 \rang}{4} \int_0^{1/B_\text{max}} \d \lambda
\frac{ \lambda}{\lang \xi \rang} \label{eq:fc}\\
f_s &= \frac{3 \lang B^2 \rang}{4} \int_0^{1/B_\text{max}} \d \lambda 
\frac{\lambda \lang g_4 \rang}{\lang \xi \rang}, \label{eq:fs}
\end{align}
are standard functions of geometry.

With this, we have that
\begin{align}
  c_1 &= \frac{f_c}{1-f_c}\label{eq:tc1},\\
  c_2 &= \frac{f_s}{1-f_c} \label{eq:tc2}, \\
  c_3 &= \frac{f_s }{1-f_c}\left[1 + f_c \left(\frac{3}{2} - \eta\right)\right] \label{eq:tc3}, \\
  c_4 &= \frac{f_c}{\alpha}\left(\frac{\{{\nu_D^{iz}}^2/\nu_D^{ii} \}}{\{\nu_{D}^{iz} \}} + \frac{f_c}{1-f_c}\frac{\{\nu_{D}^{iz}\}}{\{\nu_{D}^{ii} \}} \right) \label{eq:tc4},
\end{align}

and $K_z$ becomes
\begin{equation}
\begin{aligned}
 &K_z(\psi) \lang\frac{B^2}{n_z} \rang \\ 
= & \frac{T_i}{e} \left[f_s + \lang uB^2 \rang \right] \left(\left[\frac{f_c}{1 - f_c} + 1 \right] A_{1i} - \left[\frac{\eta f_c }{1 - f_c} + \frac{3}{2} \right] A_{2i}\right) 
- \frac{\d \lang \Phi \rang}{\d \psi} \lang \w B^2 \rang,
\end{aligned}
\end{equation}
which results in the friction-force
\begin{align}
R_{iz,\|} &\frac{\tau_{iz}}{n_i m_i} =  \left(\w  - \frac{\lang \w B^2 \rang }{n_z \lang\frac{B^2}{n_z} \rang} \right) B \frac{\d \lang \Phi \rang}{\d \psi}   \label{eq:Fiztrace}\\
&+ \left(\frac{\lang u B^2 \rang}{\lang B^2 \rang}-u + \left(\frac{1}{n_z \lang\frac{B^2}{n_z} \rang} - \frac{1}{\lang B^2 \rang}\right)  \left[f_s + \lang uB^2 \rang \right] \left[\frac{f_c}{1 - f_c} + 1 \right] \right) B \frac{T_i}{e} A_{i1} \nonumber\\
          &- \left(\frac{3}{2}\frac{\lang u B^2 \rang}{\lang B^2 \rang}-\frac{3}{2}u + \left(\frac{1}{n_z \lang\frac{B^2}{n_z} \rang} - \frac{1}{\lang B^2 \rang}\right)  \left[f_s + \lang uB^2 \rang \right] \left[\frac{\eta f_c }{1 - f_c} + \frac{3}{2} \right] \right)   B \frac{T_i}{e} A_{i2}, \nonumber
          \end{align}  

\bibliographystyle{jpp}
\bibliography{./plasma-bib} 
\end{document}